\documentclass[pra, twocolumn, floatfix, superscriptaddress, longbibliography, preprintnumbers]{revtex4-2}

\usepackage{amsmath, amssymb, bbm, bbold}
\usepackage{graphicx}
\usepackage{color}

\newcommand{\ket}[1]{{| #1 \rangle}}
\newcommand{\bra}[1]{{\langle #1 |}}

\newcommand{\Tr}{{\rm Tr}}
\newcommand{\sinc}{{\rm sinc}}

\begin{document}

\title{Unraveling the topology of dissipative quantum systems}

\author{Clemens Gneiting}
\email{clemens.gneiting@riken.jp}
\affiliation{Theoretical Quantum Physics Laboratory, RIKEN Cluster for Pioneering Research, Wako, Saitama 351-0198, Japan}
\affiliation{RIKEN Center for Quantum Computing, Wako, Saitama 351-0198, Japan}
\author{Akshay Koottandavida}
\altaffiliation{present address: Department of Applied Physics and Physics, Yale University, New Haven, CT 06520, USA}
\affiliation{Theoretical Quantum Physics Laboratory, RIKEN Cluster for Pioneering Research, Wako, Saitama 351-0198, Japan}
\author{A.V. Rozhkov}
\affiliation{Institute for Theoretical and Applied Electrodynamics, Russian Academy of Sciences, Moscow, 125412 Russia}
\author{Franco Nori}
\affiliation{Theoretical Quantum Physics Laboratory, RIKEN Cluster for Pioneering Research, Wako, Saitama 351-0198, Japan}
\affiliation{RIKEN Center for Quantum Computing, Wako, Saitama 351-0198, Japan}
\affiliation{Department of Physics, University of Michigan, Ann Arbor, Michigan 48109-1040, USA}

\date{\today}

\begin{abstract}
We discuss topology in dissipative quantum systems from the perspective of quantum trajectories. The latter emerge in the unraveling of Markovian quantum master equations and/or in continuous quantum measurements. Ensemble-averaging quantum trajectories at the occurrence of quantum jumps, i.e., the {\it jump times}, gives rise to a discrete, deterministic evolution which is highly sensitive to the presence of dark states [Gneiting~{\it et~al.}, Phys. Rev. A {\bf 104}, 062212 (2021)]. We show for a broad family of translation-invariant collapse models that the set of dark state-inducing Hamiltonians imposes a nontrivial topological structure on the space of Hamiltonians, which is also reflected by the corresponding jumptime dynamics. The topological character of the latter can then be observed, for instance, in the transport behavior, exposing an infinite hierarchy of topological phase transitions. We develop our theory for one- and two-dimensional two-band Hamiltonians, and show that the topological behavior is directly manifest for chiral, $\mathcal{PT}$, or time reversal-symmetric Hamiltonians.
\end{abstract}

\maketitle

\section{Introduction}

An elegant and powerful theory of topological phases is applicable to closed quantum systems \cite{Hasan2010colloquium, Qi2011topological, Ryu2010topological, Asboth2016short}. Conditioned on the presence of a spectral gap and generic symmetries, Hermitian Hamiltonians are classified into topological equivalence classes, labeled by topological indices that are invariant under perturbative deformations of the Hamiltonian. Bulk-boundary correspondence theorems then relate these topological invariants of the bulk to the existence of robust, gapless edge states. In topological insulators of closed systems, observing these edge states can serve to detect the underlying bulk topology.

Formulating such a theory for open, dissipative quantum systems requires new strategies, regarding both how to identify the topology and how to detect it. Some lossy classical and/or quantum systems can be characterized by non-Hermitian Hamiltonians, which allow for a spectral analysis similar to that in closed systems. While this has led to interesting insights (e.g., \cite{Rudner2009topological, Malzard2015topologically, Leykam2017edge, Venuti2017topological, Gong2017zeno, Yao2018edge, Kunst2018biorthogonal, Gong2018topological, Wang2019nonhermitian, Kawabata2019symmetry, Borgnia2020nonhermitian, Ashida2020nonhermitian}), non-Hermitian Hamiltonians capture the dynamics of dissipative quantum systems, which are usually described by (Markovian) Lindblad master equations, only in the short-time or in the Zeno limit, i.e., before the occurrence of quantum jumps.

Open quantum systems are not characterized by their Hamiltonian alone, but in addition involve Lindblad operators, which describe the effect of the environment and/or a continuous measurement. The spectral analysis of Hamiltonians can then, for instance, be replaced by a spectral analysis of the (in general mixed) steady states, which again constitute Hermitian operators. In the case of quadratic/Gaussian master equations with vanishing Hamiltonian, this has led to a successful classification of topological states, with the bulk-boundary correspondence replaced by the existence of nonlocal decoherence-free subspaces \cite{Bardyn2013topology, Bardyn2018probing}. Alternatively, the full ``Liouvillian'' (comprising Hamiltonian and Lindblad operators) can be subjected to a spectral analysis, which, again for quadratic master equations, established a topological classification associated with gapless edge states with finite lifetime \cite{Lieu2020tenfold} (see also, e.g., \cite{Dangel2018topological, Minganti2019quantum, Fei2019nonhermitian, Yoshida2020fate}).

Here, we propose a substantially different approach, based on quantum trajectories. Every Lindblad master equation can be {\it unraveled} into stochastically evolving quantum trajectories, such that their ensemble average recovers the (time-dependent) solution of the master equation. We focus here on quantum jump unravelings, where the stochastic jump events occur at discrete times. Besides their formal relation to quantum master equations, quantum trajectories can also be realized in continuous monitoring schemes, endowing the trajectories with independent physical relevance.

Instead of ensemble averaging quantum trajectories at given {\it wall times}, which would lead us back to our starting point, the Lindblad equation, we here focus on their averaging at given jump counts, i.e., {\it jump times}. This was recently established \cite{Gneiting2020jumptime} as an alternative operationally implementable way to address the collective behavior of quantum trajectories. The resulting jumptime evolution equation will serve as the basis of our analysis.

This shift of perspective offers yet another, different angle on topology in dissipative quantum systems: The jumptime evolution, which relies on the persistence of quantum jumps, is highly sensitive to the presence of dark states, where quantum jumps cease to occur. As we show, for a broad family of translation-invariant collapse models, this restates a topological classification within the space of Hamiltonians;  now, however, with the spectral-gap condition of closed systems replaced by the requirement to avoid dark state-inducing Hamiltonians. In the one- and two-dimensional two-band models considered here, this introduces winding numbers as topological invariants that substantially impact the jumptime evolution. In particular, under generic symmetry constraints on the Hamiltonian, these topological indices directly control, and thus are directly detectable (e.g., by continuous monitoring), in the transport behavior described by the jumptime evolution.

In this paper, we elaborate this theory for one- and two-dimensional two-band models, extended by collapse operators with different topological properties. On the one hand, this will lead us to introduce a characteristic phase, in analogy to the Berry phase of closed quantum systems. On the other hand, it will clarify how the topology of full-fledged dissipative quantum systems requires a description beyond (non-Hermitian) effective Hamiltonians.

\section{Jumptime unraveling}

It is instructive to introduce quantum trajectories through continuous quantum measurements, where they are physically realized and quantum jumps are detected. We restrict ourselves here to pure initial states and pure-state quantum trajectories. A continuously monitored quantum system is governed by a stochastic Schr\"odinger equation \cite{Belavkin1990stochastic, Carmichael1993open, Plenio1998quantum, Wiseman2009quantum},
\begin{align} \label{Eq:stochastic_Schroedinger_equation}
d\ket{\psi_t} = & -\frac{i}{\hbar} \left( \hat{H}_{\rm eff} + i \hbar \frac{\gamma}{2} \sum_{j \in \mathcal{I}} \bra{\psi_t} \hat{L}_j^\dagger \hat{L}_j \ket{\psi_t} \right) \ket{\psi_t} dt \\
& + \sum_{j \in \mathcal{I}} \left( \frac{\hat{L}_j \ket{\psi_t}}{\bra{\psi_t} \hat{L}_j^\dagger \hat{L}_j \ket{\psi_t}} - \ket{\psi_t} \right) dN_j(t) , \nonumber
\end{align}
where the Lindblad or jump operators $\hat{L}_j$ characterize the continuous measurement. The discrete random variables $dN_j(t) \in \{0,1\}$ describe the stochastic occurrence of the quantum jumps (recorded as ``clicks'' in the detector). They follow the statistics $\mathbb{E}_{\ket{\psi_t}}[dN_j(t)] = \gamma \bra{\psi_t} \hat{L}_j^\dagger \hat{L}_j \ket{\psi_t} dt$, and $dN_j(t) dN_k(t) = \delta_{j k} dN_j(t)$, where $\mathbb{E}_{\ket{\psi_t}}$ denotes the ensemble average over all trajectories that are in state $\ket{\psi_t}$ at the time $t$. The effective Hamiltonian $\hat{H}_{\rm eff} = \hat{H} - i \hbar \frac{\gamma}{2} \sum_{j \in \mathcal{I}} \hat{L}_j^\dagger \hat{L}_j$ captures the non-Hermitian evolution between the quantum jumps, while the added nonlinear term ensures that the state remains normalized. Note that, without loss of generality, we exclude diffusive measurement schemes, which can always be understood as emerging from pointlike jump events that cannot be resolved. Moreover, we emphasize that an arbitrary collection of Lindblad operators can always be associated with a continuous measurement.

Individual quantum trajectories $\ket{\psi_t}$ are specified by a random sequence of jump events $j_n$ at times $t_n$, $\ket{\psi_t} = \ket{\psi_t(\{(j_n,t_n)\})}$. Such jump records have been successfully observed in various experimental platforms \cite{Nagourney1986shelved, Sauter1986observation, Bergquist1986observation, Peil1999observing, Gustavsson2006counting, Fujisawa2006bidirectional, Gleyzes2007quantum, Kubanek2009photon, Neumann2010single, Vijay2011observation, Sayrin2011real, Pla2013high, Minev2019catch, Kurzmann2019optical, Naghiloo2020heat, Chen2021quantum}. If quantum trajectories are averaged or read out at wall times $t$, the ensemble-averaged state $\rho_t = \mathbb{E}[\ket{\psi_t}\bra{\psi_t}]$ follows a Gorini-Kossakowski-Sudarshan-Lindblad master equation \cite{Gorini1976completely, Lindblad1976generators},
\begin{align} \label{Eq:Lindblad_master_equation}
\partial_t \rho_t = -\frac{i}{\hbar} [\hat{H}, \rho_t] + \gamma \sum_{j \in \mathcal{I}} \Big( \hat{L}_j \rho_t \hat{L}_j^\dagger - \frac{1}{2} \{ \hat{L}_j^\dagger \hat{L}_j, \rho_t \} \Big) ,
\end{align}
where $\{\hat{A}, \hat{B}\} = \hat{A} \hat{B} + \hat{B} \hat{A}$. Averaging here effectively results in discarding the jump records, while the quantum trajectories are said to {\it unravel} the quantum master equation (\ref{Eq:Lindblad_master_equation}). Note that ensemble averages over classical noise or disorder realizations \cite{Vega2017dynamics, Gneiting2020disorder}, which also reproduce quantum master equations but are not related to continuous quantum measurements, are not considered here.

Alternatively, quantum trajectories can be ensemble-averaged or read out at given jump counts $n$, i.e., at the respective (from trajectory to trajectory varying) jumptimes $t_n$. The resulting ensemble-averaged state $\rho_n = \mathbb{E}[\ket{\psi_{t_n}}\bra{\psi_{t_n}}]$ is then governed by the discrete, deterministic jumptime evolution equation \cite{Gneiting2020jumptime},
\begin{align} \label{Eq:jumptime_evolution_equation}
\rho_{n+1} = \int_{0}^{\infty} \!\!\! \gamma d\tau \sum_{j \in \mathcal{I}} \hat{L}_j e^{-\frac{i}{\hbar} \hat{H}_{\rm eff} \tau} \rho_n e^{\frac{i}{\hbar} \hat{H}_{\rm eff}^\dagger \tau} \hat{L}_j^\dagger ,
\end{align}
which encodes the evolution of the state from one quantum jump to the next. These jumptime dynamics, which we also refer to as {\it jumptime unraveling}, lay the ground for our analysis. We emphasize that the evolution (\ref{Eq:jumptime_evolution_equation}) is exact and operationally accessible. A detailed derivation and discussion of (\ref{Eq:jumptime_evolution_equation}) are given in \cite{Gneiting2020jumptime}.

Bound to the occurrence of quantum jumps, the jumptime evolution (\ref{Eq:jumptime_evolution_equation}) describes a trace-preserving quantum map only if the dynamics do not allow for dark states \cite{Gneiting2020jumptime}. Dark states $\ket{\psi_{\rm D}}$ are pure states that satisfy $\hat{L}_j \ket{\psi_{\rm D}} = 0 \, \forall j$ and $[\hat{H}, \ket{\psi_{\rm D}}\bra{\psi_{\rm D}}] = 0$. Once the system is in a dark state, the quantum jumps cease to occur. As we show below, this sensitivity to the presence of dark states may imprint topological properties on the jumptime dynamics, even if the latter remain dark-state free.

To illustrate the jumptime evolution, let us examine an instructive example. Consider a single particle propagating on a nearest-neighbor hopping chain with hopping constant $J$ and lattice constant $a$, $\hat{H} = J \sum_{j \in \mathbb{Z}} (\ket{j+1}\bra{j}+{\rm H.c.}) = 2 J \cos \frac{\hat{p} a}{\hbar}$, extended by a dissipative directional hopping $\hat{L} = \sum_{j \in \mathbb{Z}} \ket{j+1}\bra{j} = e^{-i \hat{p} a/\hbar}$, where $\hat{p} = \int_{-h/2 a}^{h/2 a} dp \, p \, \ket{p}\bra{p}$ and $\bra{j} p \rangle = e^{i j a p/\hbar}/\sqrt{2 \pi \hbar}$. This (translation-invariant) dissipator, which translates each state by one unit cell, excludes the existence of dark states, independently from the Hamiltonian (as $\hat{L} \ket{\psi} \neq 0 \,\, \forall \ket{\psi}$). The resulting master equation (\ref{Eq:Lindblad_master_equation}) is easily solved in the momentum representation. For a localized initial state, $\rho_0 = \ket{j_0}\bra{j_0}$, we derive that the expectation value of the displacement operator $\hat{x} = \sum_{j \in \mathbb{Z}} a j \ket{j}\bra{j}$ is $\langle \hat{x} \rangle_t = a j_0 + a \gamma t$, describing a positive current. The corresponding jumptime evolution (\ref{Eq:jumptime_evolution_equation}) can also be solved analytically, yielding the respective average displacement $\langle \hat{x} \rangle_n = a j_0 + a n$. Note that the jumptime transport is independent from the dissipation rate $\gamma$, which is absorbed by the waiting time between the quantum jumps. The same transport behavior is obtained if the nearest-neighbor hopping is replaced by a general $H(\hat{p})$. In this example, the current is purely dissipation-induced, while the coherent hopping merely disperses the wave packet.

\section{Dark state-induced topology}

\begin{figure}[t]
	\includegraphics[width=0.99\columnwidth]{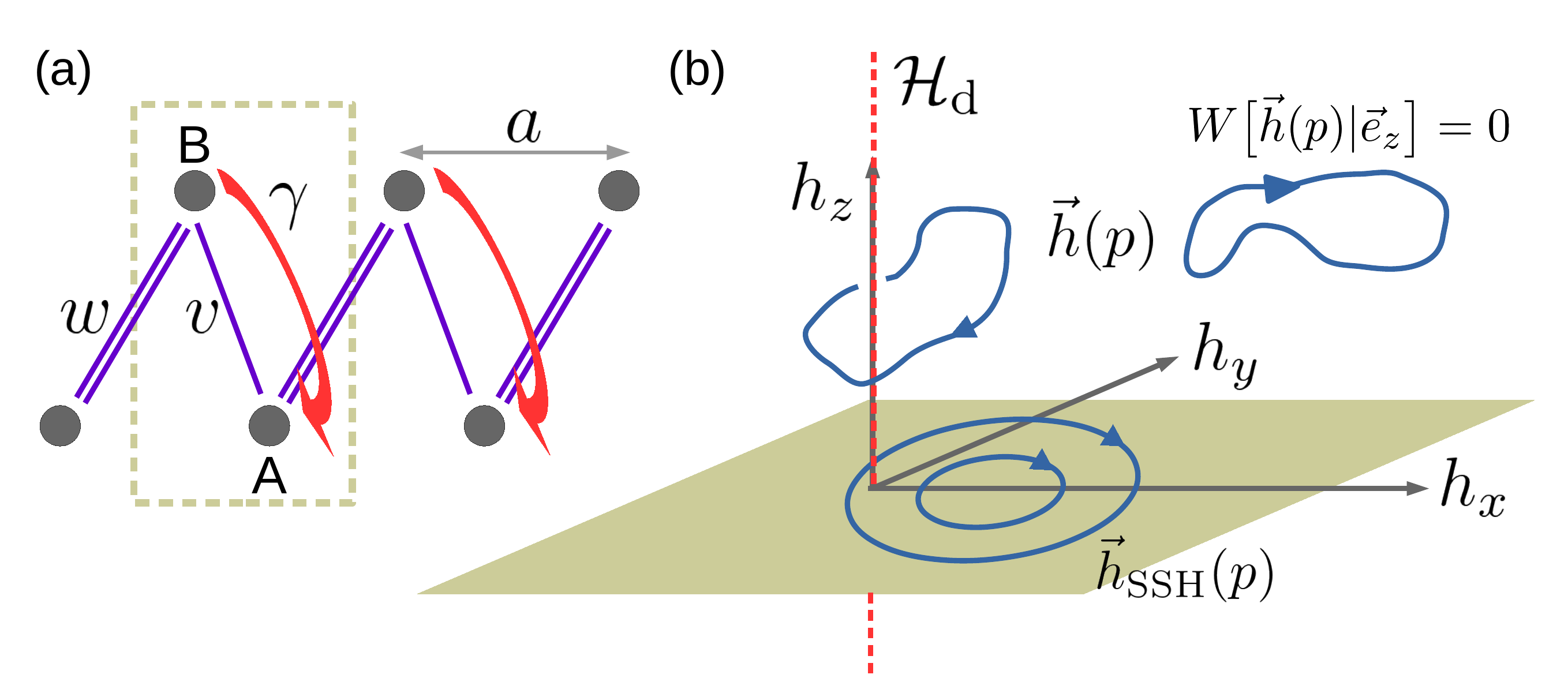}
	\caption{\label{Fig:dissipative_topology} Dark state-induced topology in one-dimensional two-band models. (a) Dissipative extension of the Su-Schrieffer-Heeger (SSH) model (beige dashed box comprising a unit cell): A collapse process from the $B$ sublattice to the $A$ sublattice (red arrows) turns states on the $A$ sublattice into dark states whenever intracell hopping $v$ and intercell hopping $w$ vanish, $v=w=0$. (b) The general set of dark state-inducing Hamiltonians $\mathcal{H}_{\rm d}$ (red dashed line) for this collapse process coincides with the $z$ axis of the Bloch space. Bloch-space Hamiltonians $\vec{h}(p)$ can then be topologically classified according to their winding number $W \big[ \vec{h}(p)|\vec{e}_z \big]$ about $\mathcal{H}_{\rm d}$.}
\end{figure}

A generic situation where translation-invariant collapse models can give rise to dark states is that of single-particle two-band models, i.e., lattices with two sites per unit cell. We first discuss one-dimensional lattices; below, we then generalize our theory to two dimensions. We thus focus now on translation-invariant Hamiltonians $\hat{H} = \oint dp \, \ket{p}\bra{p} \otimes \hat{H}(p)$ (from here on, we abbreviate the Brillouin zone integral $\int_{-h/2 a}^{h/2 a} dp \equiv \oint dp$), with the Bloch Hamiltonian
\begin{align} \label{Eq:Bloch_Hamiltonian}
\hat{H}(p) = h_x(p) \sigma_x + h_y(p) \sigma_y + h_z(p) \sigma_z .
\end{align}
The Pauli matrices $\sigma_i$, $i \in \{ x, y, z \}$, are expressed with respect to the intracell basis $\{ \ket{A}, \ket{B} \}$, using the convention $\sigma_z = \ket{A}\bra{A} - \ket{B}\bra{B}$. We remark that an additional contribution $h_0(p) \mathbb{1}_2$ would have no effect on the jumptime evolution, as is easily seen by inspecting (\ref{Eq:jumptime_evolution_equation}).

The Hamiltonian (\ref{Eq:Bloch_Hamiltonian}) includes the paradigmatic Su-Schrieffer-Heeger (SSH) model \cite{Su1979solitons} as a special case, which we sometimes use for illustration in the remainder: $h_x(p) = v + w \cos \frac{p a}{\hbar}$, $h_y(p) = -w \sin \frac{p a}{\hbar}$, and $h_z(p) \equiv 0$, where $v>0$ ($w>0$) denotes the intracell (intercell) hopping and $a$ denotes the lattice constant, cf.~Fig.~\ref{Fig:dissipative_topology}.

As a paradigmatic dissipative extension entailing dark states, we consider {\it collective collapse}, described by a single jump operator
\begin{align} \label{Eq:collective_collapse_operator}
\hat{L}_{\rm cc} = \mathbb{1}_{\infty} \otimes \ket{A}\bra{B} ,
\end{align}
where $\mathbb{1}_{\infty}$ denotes the identity in the infinite-dimensional external lattice space. In the intracell space, the jump operator induces a (momentum-independent) incoherent hopping from the $B$ to the $A$ site. While such simultaneous collapse over the extent of the entire lattice may appear artificial, it serves well for our demonstrational purposes. Below, we also discuss the case of localized collapse operators. Alternatively, replacing the identity $\mathbb{1}_{\infty}$ by $e^{-i \hat{p} a/\hbar}$, as in the example above, results in an overall drift on top. We stress that the resulting (walltime) master equations in all these cases are not quadratic/Gaussian.

Clearly, the collective collapse operator $\hat{L}_{\rm cc}$ admits the possibility of dark states, as it annihilates any state that lives exclusively on the $A$ sublattice, $\hat{L}_{\rm cc} \oint dp \, \psi(p) \ket{p} \otimes \ket{A} = 0$ for general $\psi(p)$. Whether dark states exist or not is then ultimately decided by the Hamiltonian. Specifically, $[\hat{H}(p_0), \ket{A}\bra{A}] = 0$ if and only if $\hat{H}(p_0) = h_z(p_0) \sigma_z$; that is, dark states are present whenever there exist momenta $p_0$ where $\hat{H}(p_0) = h_z(p_0) \sigma_z$. This implies that, for the dissipator $\hat{L}_{\rm cc}$, the set of dark state-inducing (or {\it dark} for short) Hamiltonians $\mathcal{H}_{\rm d}$ is determined by $\mathcal{H}_{\rm d} = \{ \hat{H} | \hat{H} = h_z \sigma_z, h_z \in \mathbb{R} \}$, which exactly comprises the $z$ axis of the Bloch space of all Hamiltonians (\ref{Eq:Bloch_Hamiltonian}), cf.~Fig.~\ref{Fig:dissipative_topology}. We thus find that, if we exclude $\mathcal{H}_{\rm d}$ from the set of admissible Bloch Hamiltonians, the resulting space $\mathcal{H}_{\rm cc}$ is not simply connected.

Due to the topology of the Brillouin zone, the Hamiltonian (\ref{Eq:Bloch_Hamiltonian}) describes a closed loop in the three-dimensional Bloch space. The topological structure of $\mathcal{H}_{\rm cc}$ then classifies the Hamiltonians (\ref{Eq:Bloch_Hamiltonian}) into separate equivalence classes, indexed by their winding number
\begin{align} \label{Eq:winding_number}
W \big[ \vec{h}(p) \big| \vec{e}_z \big] := & \nonumber \\
\oint \frac{d p}{2 \pi} \, \frac{1}{h_\perp^2(p)} & \left( \frac{\partial h_x(p)}{\partial p} h_y(p) - h_x(p) \frac{\partial h_y(p)}{\partial p} \right) ,
\end{align}
which counts the number of times the loop winds around the Bloch space's $z$ axis, the latter representing the one-dimensional manifold of dark Hamiltonians $\mathcal{H}_{\rm d}$. In Eq.~(\ref{Eq:winding_number}), $h_\perp^2(p) := h_x^2(p) + h_y^2(p)$ indicates the separation from the dark Hamiltonians (i.e., the $z$ axis). As we show below, this integer-valued topological index of the Bloch Hamiltonian can have a strong impact on the jumptime dynamics. We remark that a similar dark state-based characterization of topology has been formulated in \cite{Rudner2016survival}, there in the context of non-Hermitian/lossy quantum systems. Moreover, we note that the situation becomes more complex in the case of momentum-dependent intracell collapse, where the set of dark state-inducing Hamiltonians may in general become momentum dependent.

In the case of the SSH model, the Hamiltonian traces a circle in the $x$-$y$ plane of the Bloch space, and the winding number can take the two values $W \big[ \vec{h}(p) \big| \vec{e}_z \big] \in \{0,1\}$, with the topological transition at $v=w$, cf.~Fig.~\ref{Fig:dissipative_topology}.

\section{Topology in jumptime evolution}

We now demonstrate how the dark state-induced topological index of the Hamiltonian controls the jumptime dynamics under collective collapse (\ref{Eq:collective_collapse_operator}), even under the condition that dark Hamiltonians are avoided. Clearly, the latter is necessary in order to obtain a persistent jumptime evolution: For instance, in the extreme case of only dark Hamiltonians, $\hat{H}(p) = h_z(p) \sigma_z \, \forall p$, the jumptime evolution quickly terminates completely, $\rho_{n} = 0$ for $n>1$. Let us consider the case $h_\perp(p) \neq 0 \, \forall p$, i.e., dark Hamiltonians are avoided entirely. For clarity, we first fix $h_z(p) \equiv 0$. If we evaluate (\ref{Eq:jumptime_evolution_equation}) in the momentum basis, we obtain
\begin{align}
\bra{p} \rho_{n+1}^{(A)} \ket{p'} = K_{\rm cc}(p,p') \bra{p} \rho_{n}^{(A)} \ket{p'} ,
\end{align}
where the jumptime propagator reads
\begin{align} \label{Eq:collective_collapse_propagator}
K_{\rm cc}(p,p') = \frac{2 \hbar^2 \gamma^2 \big( h_x(p) + i h_y(p) \big) \big( h_x(p') - i h_y(p') \big)}{2 \big( h_\perp^2(p) - h_\perp^2(p') \big)^2 + \hbar^2 \gamma^2 \big( h_\perp^2(p) + h_\perp^2(p') \big)} .
\end{align}
Note that we restrict, without loss of generality, the quantum state to the $A$ sublattice, $\bra{p} \rho_{n}^{(A)} \ket{p'} := \bra{p, A} \rho_{n} \ket{p', A}$. This is possible because the collapse operator $\hat{L}_{\rm cc}$ projects the state back onto the $A$ sublattice, constraining the jumptime state $\rho_n$ to the $A$ sublattice from the first jump on. For the initial state $\rho_0$, we assume that it resides on the $A$ sublattice. The explicit solution of the jumptime evolution reads
\begin{align} \label{Eq:collective_collapse_explicit_solution}
\bra{p} \rho_{n}^{(A)} \ket{p'} = K_{\rm cc}(p,p')^n \bra{p} \rho_{0}^{(A)} \ket{p'} .
\end{align}
Since $K_{\rm cc}(p,p) = 1$, the momentum distribution is independent of the jump count $n$: $\bra{p} \rho_{n}^{(A)} \ket{p} = \bra{p} \rho_{0}^{(A)} \ket{p}$ if $h_\perp(p) \neq 0 \, \forall p$.

In order to see how the topological equivalence class of the Hamiltonian underlies the propagator (\ref{Eq:collective_collapse_propagator}), we define the phase
\begin{align} \label{Eq:jumptime_scalar}
T_{\rm cc} \big[ \vec{h}(p) \big] := \frac{i}{2 \pi} \oint dp \left[ \frac{\partial}{\partial p} K_{\rm cc}(p,p') \right]_{p'=p} .
\end{align}
The latter can be interpreted as a closed-loop integral over the connection $J_{\rm cc}(p) = i \left[ \frac{\partial}{\partial p} K_{\rm cc}(p,p') \right]_{p'=p}$, in analogy to the Berry phase and the Berry connection in closed quantum systems. Evaluating $T_{\rm cc} \big[ \vec{h}(p) \big]$ for (\ref{Eq:collective_collapse_propagator}), we obtain
\begin{align} \label{Eq:topological_jumptime_phase}
T_{\rm cc} \big[ \vec{h}(p) \big] = W \big[ \vec{h}(p) \big| \vec{e}_z \big] ,
\end{align}
i.e., $T_{\rm cc} \big[ \vec{h}(p) \big]$ coincides with the winding number about the dark Hamiltonians $\mathcal{H}_{\rm d}$, cf.~Eq.~(\ref{Eq:winding_number}). We thus take the jumptime phase (\ref{Eq:jumptime_scalar}) as an indicator for the impact of the dark-state induced topology on the jumptime evolution. Below, we show that $T_{\rm cc} \big[ \vec{h}(p) \big]$, for instance, directly controls the transport behavior.

For the more general case $h_z(p) \neq 0$, the jumptime propagator is derived in Appendix~A. A finite $h_z$ induces nontopological contributions $R_{1,2}$ to the jumptime phase:
\begin{align} \label{Eq:topological_jumptime_phase_without_symmetries}
T_{\rm cc} \big[ \vec{h}(p) \big] = W \big[ \vec{h}(p) \big| \vec{e}_z \big] + R_1 \big[ \vec{h}(p) \big] + R_2 \big[ \vec{h}(p) \big] ,
\end{align}
with
\begin{subequations} \label{Eq:rest_terms}
\begin{align}
R_1 \big[ \vec{h}(p) \big] &= -\frac{2}{\hbar \gamma} \oint \frac{dp}{2 \pi} \frac{\partial h_z(p)}{\partial p} \ln \left[ \frac{h_\perp^2(p)}{\hbar^2 \gamma^2} \right] , \\
R_2 \big[ \vec{h}(p) \big] &= \oint \frac{dp}{2 \pi} \frac{\partial h_z(p)}{\partial p} \frac{16 h_z^2(p) + \hbar^2 \gamma^2}{4 \hbar \gamma \, h_\perp^2(p)} .
\end{align}
\end{subequations}
Imposing symmetry constraints on the system Hamiltonian may, however, cause these residual terms to vanish. This holds for chiral (sublattice) symmetry or $\mathcal{PT}$ symmetry \cite{Asboth2016short}, which enforce $h_z(p) \equiv 0$, the case discussed above. Similarly, time-reversal symmetry $\mathcal{T}$ implies
\begin{align}
h_z(-p) = h_z(p) \hspace{2mm} {\rm and} \hspace{2mm} h_\perp(-p) = h_\perp(p) ,
\end{align}
which again results in vanishing nontopological terms (\ref{Eq:rest_terms}); for details, see Appendix B. We thus find that the topological character of the jumptime evolution under collective collapse (\ref{Eq:collective_collapse_operator}) is manifest for large, generic symmetry classes of Hamiltonians. In the general case, a transition between different topological equivalence classes is accompanied by a discontinuous jump of $T_{\rm cc} \big[ \vec{h}(p) \big]$, quantified by the change in the winding number.

Note that whether and how the dark state-induced topology carries over to the jumptime dynamics are not decided by the winding about the set of dark Hamiltonians $\mathcal{H}_{\rm d}$ exclusively. For instance, if we replace $\hat{L}_{\rm cc}$ by $\hat{L}'_{\rm cc} = \mathbb{1}_{\infty} \otimes \ket{B}\bra{A}$, $\mathcal{H}_{\rm d}$ is the same, but $T_{\rm cc}' \big[ \vec{h}(p) \big] = -W \big[ \vec{h}(p) \big| \vec{e}_z \big] +$\,({\rm nontopological~terms}), with the propagation restricted to the $B$ sublattice. Alternatively, we can replace $\hat{L}_{\rm cc}$ by a sublattice projection, either
\begin{align}
\hat{L}_A = \mathbb{1}_{\infty} \otimes \ket{A}\bra{A} \hspace{2mm} {\rm or} \hspace{2mm} \hat{L}_B = \mathbb{1}_{\infty} \otimes \ket{B}\bra{B} .
\end{align}
Both dissipators share with $\hat{L}_{\rm cc}$ the same set of dark Hamiltonians. However, their respective jumptime phases (which are defined analogously to (\ref{Eq:jumptime_scalar}), with $K_{\rm cc}(p,p')$ replaced by $K_A(p,p')$ and $K_B(p,p')$, respectively; $K_A(p,p')$ and $K_B(p,p')$ can be found in Appendix~C) vanish,
\begin{align} \label{Eq:jumptime_phase_sublattice_projector}
T_A \big[ \vec{h}(p) \big] = T_B \big[ \vec{h}(p) \big] = 0
\end{align}
for any $\vec{h}(p)$. This is because, even at the dark Hamiltonians, the jumptime evolution persists within the projected (i.e., nondark) sublattices; e.g., for $\hat{L}_A$, we have $\bra{p} \rho_{n+1}^{(A)} \ket{p'} = \hbar \gamma \big[ \hbar \gamma + i \big( h_z(p) - h_z(p') \big) \big]^{-1} \bra{p} \rho_{n}^{(A)} \ket{p'}$ if $\hat{H}(p) = h_z(p) \sigma_z$ $\forall p$. The jumptime propagator for $\hat{L}_A$ with $h_\perp(p) \neq 0 \, \forall p$ is given in Appendix~C. The capacity of the dissipator to shuffle states from the nondark sector of the Hilbert space to the (under dark Hamiltonians) dark sector (from the $B$ to the $A$ sublattice in the case of $\hat{L}_{\rm cc}$) is thus yet another essential prerequisite. This implies that the non-Hermitian Hamiltonian alone cannot explain the emergence of nontrivial topological jumptime behavior, as $\hat{L}_{\rm cc}$ and $\hat{L}_B$ give rise to the same
\begin{align} \label{Eq:effective_Hamiltonian}
\hat{H}_{\rm eff} = \hat{H} - i \hbar \frac{\gamma}{4} \mathbb{1}_\infty \otimes (\mathbb{1}_2 - \sigma_z) .
\end{align}
Only the jumptime propagator and jumptime phase contain the complete information to assess the impact of the dark state-induced topology.

\section{Topological transport}

A remarkable and useful property of the jumptime phase (\ref{Eq:jumptime_scalar}) is that it is directly observable in the transport behavior. To see this, we evaluate the average displacement $\langle \hat{x} \rangle_n$ for collective collapse (\ref{Eq:collective_collapse_operator}), with $\hat{x} = \sum_{j \in \mathbb{Z}} a j \ket{j}\bra{j} \otimes \mathbb{1}_2$. We can write
\begin{align}
\langle \hat{x} \rangle_n = i \hbar \oint d p \oint d p' \delta(p-p') \frac{\partial}{\partial p} \bra{p} \rho_n^{(A)} \ket{p'} .
\end{align}
Assuming a localized initial state, $\rho_0 = \ket{j_0}\bra{j_0} \otimes \ket{A}\bra{A}$ and hence $\bra{p} \rho_{0}^{(A)} \ket{p} = a/h$, one derives from (\ref{Eq:collective_collapse_explicit_solution})
\begin{align}
\langle \hat{x} \rangle_n = \langle \hat{x} \rangle_0 + i \frac{a}{2 \pi} n \oint d p \oint d p' \delta(p-p') \frac{\partial}{\partial p} K_{\rm cc}(p,p') ,
\end{align}
where we used that $K_{\rm cc}(p,p)^{n-1} = 1$. It follows immediately that
\begin{align} \label{Eq:topological_average_displacement}
\langle \hat{x} \rangle_n = a j_0 + a n T_{\rm cc} \big[ \vec{h}(p) \big] ,
\end{align}
i.e., the average displacement in jump time is, under generic symmetry constraints, controlled by the winding about the dark Hamiltonians, cf.~(\ref{Eq:topological_jumptime_phase}), changing linearly with the jump count $n$ in the topologically nontrivial sectors. In the absence of symmetries, the jumptime phase is given by (\ref{Eq:topological_jumptime_phase_without_symmetries}).

Note that, while the transport behavior (\ref{Eq:topological_average_displacement}) may be reminiscent of topological pumping \cite{Asboth2016short, Xiao2010berry}, no periodic modulation of the Hamiltonian is involved here. Rather, one may argue that the role of the latter is here taken by the (stochastically) concatenating quantum jumps.

We emphasize that (\ref{Eq:topological_average_displacement}) is valid at any stage of the jumptime evolution, from the localized initial state to the fully dispersed asymptotic state. More generally, it holds whenever the momentum distribution is homogeneous. Since the momentum distribution is a constant of motion under $\hat{L}_{\rm cc}$, this makes conceivable the rapid switching between different topological transport behaviors. In contrast, the steady state of the respective walltime master equation (\ref{Eq:Lindblad_master_equation}) does not reproduce the topological transport behavior, as shown in Appendix~D.

\begin{figure}[htb]
	\includegraphics[width=0.99\columnwidth]{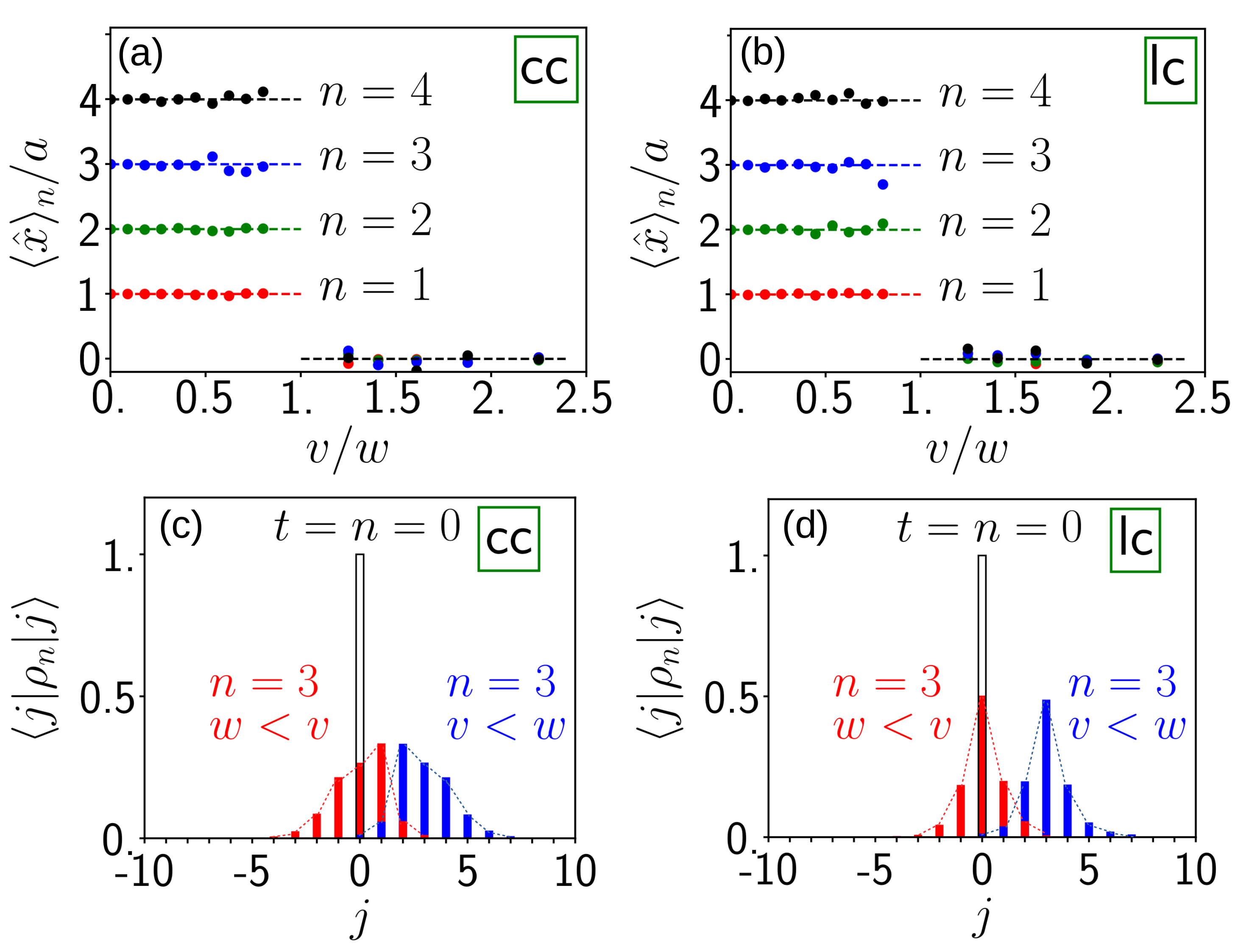}
	\caption{\label{Fig:topological_transport}
	Topological transport and wave packet evolution in jumptime unraveling for the dissipative SSH model. Both the (a) collective collapse (cc) (\ref{Eq:collective_collapse_operator}) and the (b) local collapse (lc) (\ref{Eq:local_collapse_operator}) give rise to the same topological transition: In the topologically nontrivial ($w>v$) phase, the average displacement $\langle \hat{x} \rangle_n$ grows linearly with the jump count $n$, while $\langle \hat{x} \rangle_n \equiv 0$ in the trivial ($v>w$) phase (dashed lines show analytical predictions~(\ref{Eq:topological_average_displacement}) [or, equivalently, (\ref{Eq:incremental_topological_average_displacement})], and colored dots represent numerically exact averages over $N=700$ quantum trajectories, for the initial state $\rho_0 = \ket{j_0}\bra{j_0} \otimes \ket{A}\bra{A}$ with $j_0 = 0$). At the same time, the underlying spatial distributions of the jumptime-averaged states are strongly dependent on the collapse model: Already at $n=3$, the spatially asymmetric wave packets for collective collapse (c) differ significantly from the virtually symmetric wave packets for local collapse (d). The histograms are plotted for $w=0.2 \hbar \gamma$ and $v=0.5 \hbar \gamma$ (red bars), and $w=0.5 \hbar \gamma$ and  $v=0.2 \hbar \gamma$ (blue bars). Empty bars correspond to the initial states at $n=0$.}
\end{figure}

In the case of the SSH model, we obtain $\langle \hat{x} \rangle_n = a j_0 + a n \theta(w-v)$, with the theta function $\theta(x<0)=0$, $\theta(x>0)=1$, and $\theta(x=0)=\frac{1}{2}$. In Fig.~\ref{Fig:topological_transport}, we confirm this for the first four jump counts, obtained through numerically exact ensemble averaging over $N=700$ quantum trajectories. We also depict jumptime-evolved states for the trivial and the topological phase, respectively, which highlights that the underlying topological pattern is not manifest in the spatial distributions.

A similar winding number-controlled average displacement for a non-Hermitian/lossy extension of the SSH model was derived in seminal earlier work \cite{Rudner2009topological}. In our language, this work describes the evolution up to the first jump event, restricting the average displacement to a single step (one unit cell for the SSH model). In contrast, the topological transport (\ref{Eq:topological_average_displacement}) supersedes the topological phase transition found in \cite{Rudner2009topological}, generalizing it to any jump count $n$. In monitoring implementations, this gives rise to an infinite hierarchy of topological phase transitions of arbitrary jump order $n$, where different phase transitions are operationally distinguished by the jump counts from preparation of the initial state to the concluding state read out. Note that, with increasing depth $n$, this topological pattern is increasingly hidden in the statistics of the chains of quantum jumps.

We stress again that whether a persistent jumptime current arises or not is not decided by the effective Hamiltonian alone. For instance, as elaborated in the previous section, the lattice projector $\hat{L}_B$ gives rise to the same effective Hamiltonian (\ref{Eq:effective_Hamiltonian}) as the collective collapse (\ref{Eq:collective_collapse_operator}), while its jumptime phase vanishes, cf. (\ref{Eq:jumptime_phase_sublattice_projector}), and thus no persistent jumptime transport emerges (in this case the state is constrained to the $B$ sublattice). Note that this demonstrates the divergence between our topological classification and the one presented in \cite{Rudner2009topological}, where topological behavior is predicted both for $\hat{L}_{\rm cc}$ and for $\hat{L}_B$, based on the requirement that the initial state is, independently of the jump operator, prepared on the $A$ sublattice.
	
It is worth mentioning that we can interpolate between these two cases if we consider a dissipative system that features the two jump operators
\begin{align}
\hat{L}_1 = \sqrt{\frac{\gamma_{\rm cc}}{\gamma}} \hat{L}_{\rm cc} \hspace{2mm} {\rm and} \hspace{2mm} \hat{L}_2 = \sqrt{\frac{\gamma_B}{\gamma}} \hat{L}_B ,
\end{align}
where $\gamma = \gamma_{\rm cc} + \gamma_B$. The effective Hamiltonian is again (\ref{Eq:effective_Hamiltonian}). The jumptime evolution now assumes the invariant intracell state
\begin{align}
\rho_{\rm iis} = \frac{\gamma_{\rm cc}}{\gamma} \ket{A}\bra{A} + \frac{\gamma_B}{\gamma} \ket{B}\bra{B}
\end{align}
after the first jump (or maintains it, if the intracell state of the initial state assumes this form), so that we can write $\rho_n = \rho_n^{({\rm iis})} \otimes \rho_{\rm iis}$. We then obtain
\begin{align}
\bra{p} \rho_{n+1}^{({\rm iis})} \ket{p'} = K_{{\rm cc}+B}(p,p') \bra{p} \rho_{n}^{({\rm iis})} \ket{p'} ,
\end{align}
with the scalar propagator
\begin{align} \label{Eq:combined_jumptime_propagator}
K_{{\rm cc}+B}(p,p') = \frac{\gamma_{\rm cc}}{\gamma} K_{\rm cc}(p,p') + \frac{\gamma_B}{\gamma} K_{B}(p,p') .
\end{align}
The jumptime phase with respect to (\ref{Eq:combined_jumptime_propagator}) is given by
\begin{align}
T_{{\rm cc}+B} \big[ \vec{h}(p) \big] = \frac{\gamma_{\rm cc}}{\gamma} T_{\rm cc} \big[ \vec{h}(p) \big] + \frac{\gamma_B}{\gamma} T_{B} \big[ \vec{h}(p) \big] ,
\end{align}
where $T_{B} \big[ \vec{h}(p) \big] = 0$, cf.~(\ref{Eq:jumptime_phase_sublattice_projector}). For a localized initial state, $\rho_0 = \ket{j_0}\bra{j_0} \otimes \rho_{\rm iis}$, the average displacement then evaluates as
\begin{align}
\langle \hat{x} \rangle_n = a j_0 + a n T_{{\rm cc}+B} \big[ \vec{h}(p) \big] ;
\end{align}
that is, the ratio $\gamma_{\rm cc}/\gamma$ determines the size of the displacement steps.

To see that the presence of  $\hat{L}_{\rm cc}$ does not necessarily imply topological behavior, let us consider $\hat{L}_{\rm cc}$ combined with $\hat{L}_A$, i.e., $\hat{L}_1 = \hat{L}_{\rm cc}$ and $\hat{L}_2 = \hat{L}_A$. The effective Hamiltonian then reads $\hat{H}_{\rm eff} = \hat{H} - i \hbar \frac{\gamma}{2} \mathbb{1}_\infty \otimes \mathbb{1}_2$ and the set of dark Hamiltonians is empty, i.e., there is no dark state-induced topology; correspondingly, if we evaluate the jumptime phase with respect to the propagator $\bra{p} \rho_{n+1}^{(A)} \ket{p'} = K_{{\rm cc}+A}(p,p') \bra{p} \rho_{n}^{(A)} \ket{p'}$, we obtain
\begin{align} \label{Eq:cc+A_jumptime_phase}
T_{{\rm cc}+A} \big[ \vec{h}(p) \big] = \frac{1}{2} \oint \frac{d p}{2 \pi} \, \frac{\frac{\partial h_x(p)}{\partial p} h_y(p) - h_x(p) \frac{\partial h_y(p)}{\partial p}}{h_\perp^2(p) + \hbar^2 \gamma^2/4} ,
\end{align}
where we assumed $h_z(p) \equiv 0$ for simplicity. The jumptime phase (\ref{Eq:cc+A_jumptime_phase}) does not contain a topological contribution and is, in general, nonvanishing.

\section{Localized collapse}

The collective collapse (\ref{Eq:collective_collapse_operator}) can be generalized to a broader class of translation-invariant collapse models. To this end, we introduce a collection of jump operators describing momentum kicks,
\begin{align}
\hat{L}_q = e^{i q \hat{x}/\hbar} \otimes \ket{A}\bra{B} ,
\end{align}
weighted by a momentum transfer distribution $G(q)$ with $\oint d q \, G(q) = 1$ and $G(-q) = G(q)$ \big( $\hat{x} = \sum_{j \in \mathbb{Z}} a j \ket{j}\bra{j}$ \big). The corresponding jumptime evolution (\ref{Eq:jumptime_evolution_equation}) reads
\begin{align} \label{Eq:translation-invariant_jumptime_propagation}
\rho_{n+1} = \int_{0}^{\infty} \gamma d\tau \oint d q \, G(q) \hat{L}_q e^{-\frac{i}{\hbar} \hat{H}_{\rm eff} \tau} \rho_n e^{\frac{i}{\hbar} \hat{H}_{\rm eff}^\dagger \tau} \hat{L}_q^\dagger ,
\end{align}
with both the effective Hamiltonian $\hat{H}_{\rm eff}$ [cf.~(\ref{Eq:effective_Hamiltonian})] and the set of dark Hamiltonians $\mathcal{H}_{\rm d}$ the same as for the above discussed collective collapse.

One easily verifies that the collective collapse (\ref{Eq:collective_collapse_operator}) is comprised as a limiting case, corresponding to $G(q) = \delta(q)$. The opposite limit of {\it local collapse}, where each unit cell is endowed with an individual Lindblad operator
\begin{align} \label{Eq:local_collapse_operator}
\hat{L}_j = \ket{j}\bra{j} \otimes \ket{A}\bra{B} , \hspace{5mm} j \in \mathbb{Z} ,
\end{align}
corresponds to $G(q) = a/h$. Other choices of $G(q)$ characterize localized collapse processes with spatially extended range. For instance, we can associate $G(q) \propto \exp \left[ -\sigma^2 q^2/2 \hbar^2 \right]$ with coarse-grained local collapse operators of spatial width $\sigma>a$.

The jumptime evolution for the family of collapse operators $\hat{L}_q$, together with the Hamiltonian (\ref{Eq:Bloch_Hamiltonian}) and $h_\perp(p) \neq 0 \, \forall p$, becomes
\begin{align} \label{Eq:translation-invariant_jumptime_evolution}
\bra{p} \rho_{n+1}^{(A)} \ket{p'} = \oint dq \, G(q) \, K_q(p,p') \bra{p-q} \rho_{n}^{(A)} \ket{p'-q} ,
\end{align}
where $K_q(p,p') = K_{\rm cc}(p-q,p'-q)$ with the collective collapse propagator $K_{\rm cc}(p,p')$ as in (\ref{Eq:collective_collapse_propagator}) [$h_z(p) \equiv 0$] or (\ref{Eq:general_collective_collapse_propagator}) [general $h_z(p)$], respectively. The corresponding momentum distribution $\bra{p} \rho_{n+1}^{(A)} \ket{p} = \oint dq \, G(q) \bra{p-q} \rho_{n}^{(A)} \ket{p-q}$ broadens after each jump, unless $G(q) \propto \delta(q)$ (collective collapse). This broadening eventually produces the homogeneous momentum distribution required to observe topological transport, independently from the initial state. This generally happens before the steady state is reached. For instance, for local collapse, $G(q) = a/h$, a single jumptime step induces such homogeneous momentum distribution.

To accommodate the more general class of translation-invariant collapse models (\ref{Eq:translation-invariant_jumptime_propagation}), we now define the jumptime phase as
\begin{align} \label{Eq:translation-invariant_jumptime_phase}
T_{\rm ti} \big[ \vec{h}(p) \big] := \frac{i}{2 \pi} \oint dp \oint d q \, G(q) \left[ \frac{\partial}{\partial p} K_q(p,p') \right]_{p'=p} .
\end{align}
Note that the jumptime phase (\ref{Eq:translation-invariant_jumptime_phase}) reduces to (\ref{Eq:jumptime_scalar}) for $G(q) = \delta(q)$. Irrespective of $G(q)$, we find
\begin{align} \label{Eq:invariant_jumptime_phase}
T_{\rm ti} \big[ \vec{h}(p) \big] = T_{\rm cc} \big[ \vec{h}(p) \big] ,
\end{align}
i.e., the jumptime phase is invariant under the translation-invariant generalization of the collapse model. It is clear that this immutability also holds for similar translation-invariant generalizations of, e.g., $\hat{L}_{\rm cc}'$ or the sublattice projections $\hat{L}_A$ and $\hat{L}_B$, respectively.

Given $\bra{p} \rho_{n}^{(A)} \ket{p} = a/h$, we derive from (\ref{Eq:translation-invariant_jumptime_evolution}) the average displacement
\begin{align} \label{Eq:incremental_topological_average_displacement}
\langle \hat{x} \rangle_{n+1} = \langle \hat{x} \rangle_{n} + a T_{\rm ti} \big[ \vec{h}(p) \big] ;
\end{align}
that is, with (\ref{Eq:invariant_jumptime_phase}), we find that the transport under general localized collapse remains controlled by the jumptime phase~(\ref{Eq:jumptime_scalar}), i.e., by the winding about the dark Hamiltonians. (Analogously, the jumptime phase remains vanishing for localized versions of the sublattice projections.) In Fig.~\ref{Fig:topological_transport} we numerically confirm this for the SSH model with local collapse (\ref{Eq:local_collapse_operator}) for the first four jump counts. We also numerically verified it for $\hat{L}_j = \frac{1}{2} (\ket{j}\bra{j} + \ket{j+1}\bra{j+1}) \otimes \ket{A}\bra{B}$ (not displayed).

We conclude this section with two remarks: (i) For a model with the local collapse (\ref{Eq:local_collapse_operator}), and $h_z(p) = {\rm const}$, the steady-state transport of (\ref{Eq:Lindblad_master_equation}) has been shown \cite{Kastoryano2019topological} to demonstrate the topological switching, similar to our findings for the jumptime evolution. We further conjecture that this steady-state feature may also hold for other $G(q) \neq \delta(q)$. However, there is an important distinction between the steady-state transport and the jumptime transport: The jumptime dynamics reflects the topology at any stage of the evolution (given a homogeneous momentum distribution), and it also includes the collective collapse $G(q) = \delta(q)$. (ii) In the case of more general collapse models, e.g., with momentum-dependent jump operators $\hat{L}_j(p)$, we must expect that the set of dark Hamiltonians and the invariant intracell state become momentum-dependent. In such a case, the associated jumptime propagator and jumptime phase will account for the momentum dependencies of both the Hamiltonian and the jump operators, calling for a correspondingly generalized geometric underpinning.

\section{Two-dimensional lattices}

Finally, we discuss how our theory is generalized to two-dimensional two-band models. The Hamiltonian is then given by $\hat{H}^{(2{\rm D})} = \oint dp_1 \oint dp_2 \, \ket{p_1}\bra{p_1} \otimes \ket{p_2}\bra{p_2} \otimes \hat{H}^{(2d)}(\vec{p})$, where the Bloch Hamiltonian reads
\begin{align} \label{Eq:2D_Bloch_Hamiltonian}
\hat{H}^{(2{\rm D})}(\vec{p}) = h_x(\vec{p}) \sigma_x + h_y(\vec{p}) \sigma_y + h_z(\vec{p}) \sigma_z ,
\end{align}
with $\vec{p} = (p_1, p_2)^T$ being a vector composed of the two lattice momenta $p_1$ and $p_2$. Due to the topology of the Brillouin zone, the Hamiltonian (\ref{Eq:2D_Bloch_Hamiltonian}) now generically defines a torus-like surface in the Bloch space.

We complement the Hamiltonian with a two-dimensional collective collapse,
\begin{align} \label{Eq:2d_collective_collapse_operator}
\hat{L}_{\rm cc}^{(2{\rm D})} = \mathbb{1}_{\infty} \otimes \mathbb{1}_{\infty} \otimes \ket{A}\bra{B} .
\end{align}
The set of dark Hamiltonians $\mathcal{H}_{\rm d} = \{ \hat{H} | \hat{H} = h_z \sigma_z, h_z \in \mathbb{R} \}$ then remains the same as in the one-dimensional case, coinciding with the $z$ axis of the Bloch space. This gives rise to a similar topological classification of Hamiltonians, based on the winding of the toruslike Hamiltonian about the $z$ axis. However, due to the two-dimensional extension of the Bloch Hamiltonian, the latter is now characterized by two topological indices.

We introduce the two winding numbers $(i,j \in \{ 1,2 \}, i \neq j)$
\begin{align} \label{Eq:2d_winding_number}
W_i \big[ \vec{h}(\vec{p}) \big| \vec{e}_z \big] := & \nonumber \\
\oint \frac{d p_j a_j}{h} \oint \frac{d p_i}{2 \pi} & \, \frac{1}{h_\perp^2(\vec{p})} \left( \frac{\partial h_x(\vec{p})}{\partial p_i} h_y(\vec{p}) - h_x(\vec{p}) \frac{\partial h_y(\vec{p})}{\partial p_i} \right) ,
\end{align}
associating separate topological indices with the two lattice momenta. Note that we assign different lattice constants $a_j$ to the conjugate spatial coordinates. Here, the integral over $p_j$ has no effect on the outcome, i.e., the same winding number emerges for any value of $p_j$. As in the one-dimensional case, only Hamiltonians with $h_\perp(\vec{p}) \neq 0 \, \forall p_1, p_2$ (i.e., Hamiltonians that avoid dark states) admit a topological classification.

The resulting jumptime evolution (which is again restricted to the $A$ sublattice),
\begin{align}
\bra{\vec{p}} \rho_{n+1}^{(A)} \ket{\vec{p}{\,'}} = K_{\rm cc}^{(2{\rm D})}(\vec{p}, \vec{p}{\,'}) \bra{\vec{p}} \rho_{n}^{(A)} \ket{\vec{p}{\,'}} ,
\end{align}
is characterized by the jumptime propagator ($h_\perp^2(\vec{p}) = h_x^2(\vec{p}) + h_y^2(\vec{p}) \neq 0 \, \forall p_1, p_2$)
\begin{subequations} \label{Eq:2d_collective_collapse_propagator}
\begin{align}
K_{\rm cc}^{(2{\rm D})}(\vec{p}, \vec{p}{\,'}) = \frac{2 \hbar^2 \gamma^2 \big( h_x(\vec{p}) + i h_y(\vec{p}) \big) \big( h_x(\vec{p}{\,'}) - i h_y(\vec{p}{\,'}) \big)}{2 A^2(\vec{p}, \vec{p}{\,'}) + \hbar^2 \gamma^2 B(\vec{p}, \vec{p}{\,'})} ,
\end{align}
where we abbreviated
\begin{align}
A(\vec{p}, \vec{p}{\,'}) =& h_\perp^2(\vec{p}) - h_\perp^2(\vec{p}{\,'}) + h_z^2(\vec{p}) - h_z^2(\vec{p}{\,'}) \\
&+ i \hbar \gamma \big[ h_z(\vec{p}) + h_z(\vec{p}{\,'}) \big]/2 , \nonumber \\
B(\vec{p}, \vec{p}{\,'}) =& h_\perp^2(\vec{p}) + h_\perp^2(\vec{p}{\,'}) + h_z^2(\vec{p}) + h_z^2(\vec{p}{\,'}) \\
&+ i \hbar \gamma \big[ h_z(\vec{p}) - h_z(\vec{p}{\,'}) \big]/2. \nonumber
\end{align}
\end{subequations}
We remark that the propagator (\ref{Eq:2d_collective_collapse_propagator}) is identical to its one-dimensional counterpart for general $h_z(p) \neq 0$, Eq. (\ref{Eq:general_collective_collapse_propagator}), with the replacement $p \rightarrow \vec{p}$.

In order to reveal the topological content of the two-dimensional jumptime evolution described by the propagator (\ref{Eq:2d_collective_collapse_propagator}), we define the two averaged jumptime phases $(i,j \in \{ 1,2 \}, i \neq j)$
\begin{align}
\overline{T}_{{\rm cc}, i} \big[ \vec{h}(\vec{p}) \big] := \oint \frac{dp_j a_j}{h} \oint \frac{dp_i}{2 \pi} J_{{\rm cc}, i}(\vec{p}) ,
\end{align}
which we express here in terms of the jumptime connection
\begin{align} \label{Eq:jumptime_connection}
\vec{J}_{\rm cc}(\vec{p}) = i \left[ \vec{\nabla}_{\vec{p}} K_{\rm cc}^{(2{\rm D})}(\vec{p}, \vec{p}{\,'}) \right]_{\vec{p}{\,'} = \vec{p}} .
\end{align}
If we evaluate the jumptime phases for the propagator (\ref{Eq:2d_collective_collapse_propagator}), we obtain
\begin{align}
\overline{T}_{{\rm cc}, i} \big[ \vec{h}(\vec{p}) \big] = W_i \big[ \vec{h}(\vec{p}) \big| \vec{e}_z \big] + \overline{R}_{1, i} \big[ \vec{h}(\vec{p}) \big] + \overline{R}_{2, i} \big[ \vec{h}(\vec{p}) \big] .
\end{align}
The residual terms are here given by $(i,j \in \{ 1,2 \}, i \neq j)$
\begin{subequations} \label{Eq:2d_rest_terms}
\begin{align}
\overline{R}_{1, i} \big[ \vec{h}(\vec{p}) \big] &= -\frac{2}{\hbar \gamma} \oint \frac{dp_j a_j}{h} \oint \frac{dp_i}{2 \pi} \frac{\partial h_z(\vec{p})}{\partial p_i} \ln \left[ \frac{h_\perp^2(\vec{p})}{\hbar^2 \gamma^2} \right] , \\
\overline{R}_{2, i} \big[ \vec{h}(\vec{p}) \big] &= \oint \frac{dp_j a_j}{h} \oint \frac{dp_i}{2 \pi} \frac{\partial h_z(\vec{p})}{\partial p_i} \frac{16 h_z^2(\vec{p}) + \hbar^2 \gamma^2}{4 \hbar \gamma \, h_\perp^2(\vec{p})} .
\end{align}
\end{subequations}
All four residual terms vanish if $h_z(-\vec{p}) = h_z(\vec{p})$ and $h_\perp(-\vec{p}) = h_\perp(\vec{p})$, conditions which are satisfied by time reversal-symmetric two-dimensional Hamiltonians. While $h_z(\vec{p}) \equiv 0$ (corresponding to chiral or $\mathcal{PT}$ symmetry) also causes all four residual terms to vanish, the Hamiltonian is then bound to the $x$-$y$ plane of the Bloch space.

For example, let us consider a time reversal-symmetric Bloch Hamiltonian defined by ($u,v,w>0$, $v>w$)
\begin{subequations} \label{Eq:torus}
\begin{align}
h_x(\vec{p}) &= u + v \cos \left( \frac{p_1 a_1}{\hbar} \right) , \\
h_y(\vec{p}) &= v \sin \left( \frac{p_1 a_1}{\hbar} \right) + 2 w \sin \left( \frac{p_2 a_2}{\hbar} \right) , \\
h_z(\vec{p}) &= 2 w \cos \left( \frac{p_2 a_2}{\hbar} \right) ,
\end{align}
\end{subequations}
which can be considered as a two-dimensional generalization of the SSH model. Its lattice representation and Bloch-space representation are shown in Fig.~\ref{Fig:two_dimensions}. The corresponding real-space Hamiltonian is given in Appendix~E. Note that the Bloch-space manifold degenerates to a ``pinch'' line in the $x$-$z$ plane. Similar to the SSH model, the first jumptime phase experiences a topological phase transition at $u=v$, where
\begin{align}
\overline{T}_{{\rm cc}, 1} \big[ \vec{h}(\vec{p}) \big] = \Big\{
\begin{array}{r} -1, \hspace{1mm} u<v \\ 0, \hspace{1mm} u>v \end{array} .
\end{align}
In contrast, $\overline{T}_{{\rm cc}, 2} \big[ \vec{h}(\vec{p}) \big] \equiv 0$ for any parameter choice consistent with $v>w$. Let us note that, in the general case, unlike the example discussed above, the condition that the Bloch Hamiltonian does not intersect with the dark Hamiltonians may exclude an extended parameter range. For instance, in a modification of model (\ref{Eq:torus}), the parameter $u$ may be removed from the $x$ component of the Hamiltonian and added to the $y$ component, instead. In this case, variation of $u$ would shift the toruslike surface along the $y$ axis, and the intersection with the $z$ axis would occur over a finite interval of $u$ parameters.

\begin{figure}[htb]
\includegraphics[width=0.99\columnwidth]{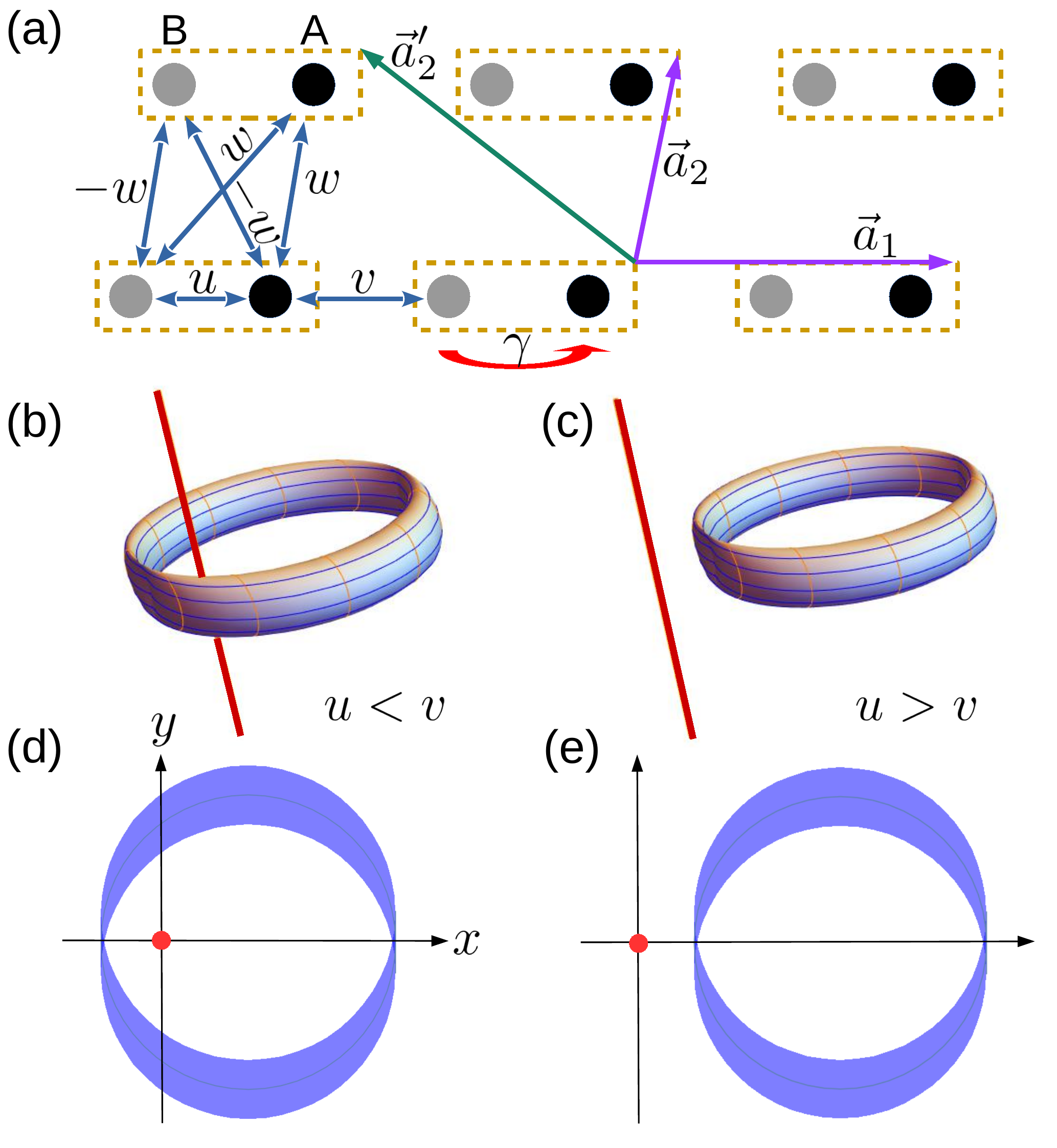}
\caption{\label{Fig:two_dimensions} Real-space (a) and Bloch-space (b-e) representation of the two-dimensional Hamiltonian~(\ref{Eq:torus}). The latter is formulated with respect to the primitive translation vectors $\vec{a}_1$ and $\vec{a}_2$, respectively. Depending on whether the set of dark state-inducing Hamiltonians (red line, coinciding with the $z$ axis of the Bloch space) is encircled (b) or not (c), the dissipative system is topologically nontrivial (b) or trivial (c). The mesh lines indicate one-dimensional loops of constant $p_1$ (blue) and $p_2$ (orange), respectively. The corresponding projections of the torus-like surfaces into the $x$-$y$ plane are shown in (d) and (e), respectively. Only the spatial coordinate $x_1$ associated with the encircling momentum $p_1$ can exhibit a nonvanishing jumptime phase and thus topological transport. The chosen parameters are $v/w=10$ for both cases, and $u/w=6$ for (b) and (d) and $u/w=14$ for (c) and (e), respectively. A different choice of primitive translation vectors, e.g., $\vec{a}_1$ and $\vec{a}'_2$, results in a transformed Bloch Hamiltonian, while the transport behavior is invariant.}
\end{figure}

As in the one-dimensional case, the topological nature of the jumptime dynamics is directly observable in the transport behavior and here extends to a topologically controlled orientation of the bulk current. This is seen by evaluating the expectation value of the position vector $\langle \hat{\vec{x}} \rangle = \langle \hat{x}_1 \rangle \vec{a}_1/a_1 + \langle \hat{x}_2 \rangle \vec{a}_2/a_2$, where $\hat{x}_1 = \sum_{j \in \mathbb{Z}} a_1 j \ket{j}\bra{j} \otimes \mathbb{1}_{\infty} \otimes \mathbb{1}_2$ and $\hat{x}_2 = \mathbb{1}_{\infty} \otimes \sum_{j \in \mathbb{Z}} a_2 j \ket{j}\bra{j} \otimes \mathbb{1}_2$ are the discrete coordinates (canonically conjugate to $p_1$ and $p_2$, respectively) that parametrize the lattice in multiples of the primitive translation vectors $\vec{a}_1$ and $\vec{a}_2$, respectively. For the initial state $\rho_0 = \ket{j_{1,0}}\bra{j_{1,0}} \otimes \ket{j_{2,0}}\bra{j_{2,0}} \otimes \ket{A}\bra{A}$, this yields ($i \in \{ 1,2 \}$)
\begin{align}
\langle \hat{x}_i \rangle_n = a_i j_{i, 0} + a_i n \overline{T}_{{\rm cc}, i} \big[ \vec{h}(\vec{p}) \big] ;
\end{align}
that is, for two-dimensional Hamiltonians that satisfy the symmetry constraints, the transport is topologically quantized. Similar to the one-dimensional case, the topology can be detected in a bulk property. It is clear that the generalization to other collapse scenarios, e.g., localized collapse models, follows the same lines as in the one-dimensional case. Moreover, it is clear that more than two spatial dimensions would result in a similar topological classification.

It is worth mentioning that the topological transport is invariant under the choice of the primitive translation vectors, even though the Bloch Hamiltonian transforms under such variation. For instance, if we rewrite a Bloch Hamiltonian $\vec{h}(\vec{p})$, which is based on the primitive translation vectors $\vec{a}_1$ and $\vec{a}_2$, with respect to the primitive translation vectors $\vec{a}'_1 = \vec{a}_1$ and $\vec{a}'_2 = \vec{a}_2 + m \vec{a}_1$ ($m \in \mathbb{Z}$), we obtain the Bloch Hamiltonian
\begin{align}
\vec{h}'(p_1, p_2) = \vec{h}(p_1, \frac{a'_2}{a_2} p_2 - m \frac{a_1}{a_2} p_1) .
\end{align}
In Appendix~E we demonstrate this for the Bloch Hamiltonian (\ref{Eq:torus}) with $m=-1$. The jumptime phases then transform as
\begin{align} \label{Eq:jumptime-phase_transformation}
\overline{T}_{{\rm cc}, 1} \big[ \vec{h}'(\vec{p}) \big] = \overline{T}_{{\rm cc}, 1} \big[ \vec{h}(\vec{p}) \big] - m \overline{T}_{{\rm cc}, 2} \big[ \vec{h}(\vec{p}) \big]
\end{align}
and $\overline{T}_{{\rm cc}, 2} \big[ \vec{h}'(\vec{p}) \big] = \overline{T}_{{\rm cc}, 2} \big[ \vec{h}(\vec{p}) \big]$, and hence
\begin{align}
\overline{T}_{{\rm cc}, 1} \big[ \vec{h}'(\vec{p}) \big] \vec{a}'_1 &+ \overline{T}_{{\rm cc}, 2} \big[ \vec{h}'(\vec{p}) \big] \vec{a}'_2 \nonumber \\
&= \overline{T}_{{\rm cc}, 1} \big[ \vec{h}(\vec{p}) \big] \vec{a}_1 + \overline{T}_{{\rm cc}, 2} \big[ \vec{h}(\vec{p}) \big] \vec{a}_2 .
\end{align}
To illustrate the use of Eq.~(\ref{Eq:jumptime-phase_transformation}) let us consider the following example. Assume that the Hamiltonian is time reversal symmetric, thus, the jumptime phases coincide with winding numbers in case of the two-dimensional collective collapse (\ref{Eq:2d_collective_collapse_operator}). If, in addition, $\overline{T}_{{\rm cc}, 1} \big[ \vec{h}(\vec{p}) \big] = s \in \mathbb{Z}$ and $\overline{T}_{{\rm cc}, 2} \big[ \vec{h}(\vec{p}) \big] = 1$, we can switch to new primitive translation vectors $\vec{a}'_1 = \vec{a}_1$ and $\vec{a}'_2 = \vec{a}_2 + s \vec{a}_1$ for which only one jumptime phase is nonvanishing. In other words, the topological transport is directed along one of the primitive translation vectors. We remark that, in cases where both jumptime phases are nonvanishing, the one-dimensional loops with either $p_1$ or $p_2$ fixed each have a toroidal component, i.e., they wind around the $z$ axis, while they can in addition exhibit a poloidal component.

As a final remark, we would like to point out that the jumptime connection (\ref{Eq:jumptime_connection}) allows us to define the curvature tensor ($i,j \in \{1,2\}$)
\begin{align}
\Omega_{{\rm cc}, ij}(\vec{p}) = \frac{\partial}{\partial p_i} J_{{\rm cc}, j}(\vec{p}) - \frac{\partial}{\partial p_j} J_{{\rm cc}, i}(\vec{p}) .
\end{align}
This is in analogy to the Berry curvature, which is defined in terms of the Berry connection. However, the corresponding Chern number, $C_{\rm cc}[\vec{h}(\vec{p})] = \oint \frac{dp_1}{2 \pi} \oint \frac{dp_2}{2 \pi} \Omega_{{\rm cc}, 12}(\vec{p})$, vanishes whenever $h_\perp(\vec{p}) \neq 0 \, \forall p_1, p_2$ (the precondition for topological classification), underscoring the different nature of the source of topology.

\section{Experimental realization}

The rapid progress in the development of quantum processing devices brings into reach the possibility of emulating the dynamics of complex quantum systems. While digital quantum simulators will require error correction, near-term analog quantum simulators (e.g., based on trapped ions or superconducting circuits) can be expected to be able to simulate one- and two-dimensional single-particle lattice models with local hoppings [e.g., the one-dimensional SSH model, or the two-dimensional model (\ref{Eq:torus})] for sufficiently long evolution times to observe the topological properties discussed here. For instance, trapped-ion systems have been used to demonstrate environment-assisted quantum transport in qubit networks \cite{Maier2019environment, Monroe2021programmable}, and superconducting qubits have been deployed to reproduce the spectrum of two-dimensional electrons in a magnetic field \cite{Roushan2017spectroscopic} (see also, e.g., \cite{Leseleuc2019observation, Lv2021measurement}).

The quantum simulation of the dynamics of single-particle models with local hoppings, while highly platform dependent, follows standard lines. We thus focus here on the implementation of the monitoring, which is required in order to observe the jumptime evolution (\ref{Eq:jumptime_evolution_equation}). For the sake of concreteness, let us focus on the implementation of a one-dimensional two-band lattice model, complemented by the monitoring of the local collapse operators (\ref{Eq:local_collapse_operator}). Moreover, let us assume that our quantum simulator consists of a one-dimensional array of qubits, restricted to the single-excitation subspace. An array of $2 N$ qubits can then serve to emulate a lattice consisting of $N$ unit cells, and the basis states that span the available Hilbert space are $\ket{1, B} = \ket{1_1, 0_2, \dots, 0_{2 N}}$, $\ket{1, A} = \ket{0_1, 1_2, 0_3, \dots, 0_{2 N}}$, $\dots$, $\ket{N, A} = \ket{0_1, \dots, 0, 1_{2 N}}$. The local collapse operators (\ref{Eq:local_collapse_operator}) are then expressed in the single-excitation subspace as
\begin{align} \label{Eq:spin-chain_local_collapse}
\hat{L}_j^{({\rm ses})} = \sigma_+^{(j, A)} \otimes \sigma_-^{(j, B)} \hspace{1mm} , \hspace{1mm} j = 1, \dots, N ,
\end{align}
where $\sigma_+^{(j, A)}$ creates an excitation at the qubit representing the $A$ site of the $j$th unit cell and $\sigma_-^{(j, B)}$ annihilates an excitation at the qubit representing the $B$ site of the $j$th unit cell.

One can show that the monitoring of the local collapse (\ref{Eq:spin-chain_local_collapse}) can be realized by complementing the qubit array with an array of ancilla qubits. Importantly, this requires only a single ancilla qubit per unit cell, and each ancilla qubit is coupled only to the qubits that comprise its associated unit cell. The system and ancilla qubits are coupled such that stroboscopic repetitions of standard projective measurements on the ancilla qubits effect the desired (quasi-)continuous monitoring on the system qubits. The repetition rate $1/\Delta t$ of the ancilla measurements is chosen such that $\Delta t$ is small compared to the typical time scales of the system dynamics.

Specifically, after each time step $\Delta t$, all ancilla qubits (which initially are prepared in their ground states) are projectively measured in their canonical basis. By design of the system-ancilla coupling, detecting the $j$th ancilla in its excited state $\ket{1^{(j,a)}}$ then indicates the occurrence of a quantum jump (\ref{Eq:spin-chain_local_collapse}) at the $j$th unit cell, while the simultaneous detection of all ancillae in their ground states $\ket{0^{(j,a)}}$ corresponds to the ``null outcome'' of no quantum jumps occurring. The probabilities for these events to occur are described by the system-level positive operator-valued measure (POVM)
\begin{subequations}
\begin{align}
\hat{F}_j &= \gamma \Delta t \hat{L}_j^{({\rm ses}) \dagger} \hat{L}_j^{({\rm ses})} \hspace{1mm} , \hspace{1mm} j = 1, \dots, N , \\
\hat{F}_0 &= \mathbb{1} - \gamma \Delta t \sum_{j=1}^N \hat{L}_j^{({\rm ses}) \dagger} \hat{L}_j^{({\rm ses})} , \label{Eq:null_outcome}
\end{align}
\end{subequations}
where $\Tr[\hat{F}_j \rho_t]$ describes the probability for a quantum jump at the $j$th unit cell, $\Tr[\hat{F}_0 \rho_t]$ represents the probability for the ``null outcome'' of no jump detections, and $\rho_t$ denotes the system state at the time of the ancilla measurements. Notice that $\Delta t$ must be chosen to be sufficiently small such that (\ref{Eq:null_outcome}) represents a positive operator. Moreover, note that, by construction, the simultaneous detection of more than one ancilla in the excited state is highly suppressed and can thus be neglected. We emphasize that,  in  the limit of sufficiently small $\Delta t$, this description is consistent with the dynamics of the quantum trajectories (\ref{Eq:stochastic_Schroedinger_equation}). After each of the stroboscopic ancilla measurements, all ancillae are reset into their ground states.

The operational procedure to detect the transport in jumptime (\ref{Eq:incremental_topological_average_displacement}) is now as follows: The initial state is prepared as a single, localized excitation in the center of the qubit array. Subsequently, the monitoring is performed through the above described system-ancilla coupling. Immediately after the detection of the desired number of quantum jumps (which is chosen such that the system state has not yet been affected by the lattice boundaries), the system qubits are all simultaneously subjected to projective readout measurements, which locates the excitation in a single unit cell. Repeating this protocol, from the preparation of the initial state to the final readout, multiple times, the average position of the excitation is then, if uncontrolled and hence detrimental environment-coupling is negligible, described by the topology-controlled transport (\ref{Eq:incremental_topological_average_displacement}).

\section{Discussion and Conclusions}

We introduced a topological classification for translation-invariant open quantum systems, based on the ensemble-averaged behavior of the associated quantum trajectories when read out at the jumptimes. The classification is rooted in the sensitivity of the jumptime evolution to the presence of dark states. Quantum trajectories that arrive at dark states cease to exhibit quantum jumps, resulting in a trace-decreasing jumptime evolution. To guarantee that each trajectory exhibits an infinite chain of quantum jumps, we thus exclude Hamiltonians that give rise to dark states. Depending on the choice of Lindblad operators, the space of nondark Hamiltonians may then acquire nontrivial connectivity, which can manifest itself as a topological contribution to the jumptime phase. Since the jumptime phase is directly related to an observable bulk property, the transport current, a hierarchy of topological phase transitions indexed by the jump order $n$ may be revealed through transport measurements.

In contrast to topological pumping in closed quantum systems, this hierarchy of topological phase transitions is not a consequence of periodic modulation, and can be also observed in the transient behavior. While the jumptime-based topological classification is not reliant on symmetry protection, i.e., it can persist in the absence of Hamiltonian symmetries, generic symmetry constraints, such as time-reversal invariance, render jumptime phase and transport entirely controlled by the topological properties of the model.

The jumptime evolution and its associated topological transport are observable in continuous monitoring schemes, which may be realized, for instance, in engineered quantum systems including trapped ions, ultracold atoms in optical lattices, or quantum simulators based on superconducting circuits.

We demonstrated the jumptime-based topological classification and the resulting topologically controlled transport with one- and two-dimensional two-band models, extended by, e.g., collective collapse $\hat{L}_{\rm cc}$, sublattice projectors $\hat{L}_A$ and $\hat{L}_B$, and generalized localized translation-invariant versions thereof, characterized by a momentum transfer distribution $G(q)$. We observe that the jumptime phase correctly discriminates the respective topological transport behaviors for these different choices of jump operators, in particular, also when these different dissipative systems share the same effective (non-Hermitian) Hamiltonian. On the other hand, the jumptime phase (and correspondingly the transport) reveals an invariance under variation of the localization pattern $G(q)$, which extends the scope of admissible topological deformations from the Hamiltonian domain into the domain of Lindblad operators.

The introduced topological classification represents a distinct alternative to other classification schemes for dissipative systems, e.g., based on (walltime) steady states. It is built on the physical requirement that all quantum trajectories exhibit an infinite chain of quantum jumps, which here takes the role of the spectral-gap requirement for closed quantum systems. Notably, the collective collapse model features a topological jumptime phase which is observable in the jumptime transport, while the respective walltime steady state fails to display topological behavior. Being based on the jumptime propagation, this topological classification is of intrinsically dynamical nature and also does not merge with closed-system classification schemes (which are based on the stationary states of Hermitian Hamiltonians), not even in the limit of small dissipation rate $\gamma$. Instead, the topological behavior revealed here holds equally for any $\gamma > 0$ and can manifestly diverge from the behavior predicted by the Hamiltonian alone, as seen, e.g., for the SSH Hamiltonian complemented by a sublattice projection (in which case the jumptime phase vanishes for any choice of the staggered hoppings).

On the other hand, by virtue of its reference to quantum trajectories, our classification connects to, can be compared with, and extends topological classifications for non-Hermitian (i.e., conditioned on no quantum jumps) quantum systems. For instance, an exhaustive classification scheme for non-Hermitian Hamiltonians has been introduced in \cite{Kawabata2019symmetry}. It is primarily based on topological properties of the spectrum when considered as a continuous set in the complex plane $\mathbb{C}$. Proposed classes of non-Hermitian Hamiltonians are based on topological connectivity characterizations of the spectrum. Specifically, spectra with nontrivial connectedness may, depending on their structure, be classified through ``point gaps'' or ``line gaps''. One finds, however, that the dark state-induced topology, encoded in our effective Hamiltonians, is not naturally captured by these categories, as the connectivity properties of the spectra of our effective Hamiltonians are not related to the topological transport. To see this, let us consider the effective Hamiltonian (\ref{Eq:effective_Hamiltonian}), with $\hat{H}$ being the SSH Hamiltonian. We then observe that, for any choice $v \neq w$ of the hopping parameters, the complex eigenvalues of the effective Hamiltonian have a positive imaginary part, i.e., they lie in the upper half of the complex plane. On the other hand, at the transition, i.e., when $v = w$ and the Hamiltonian is dark state-inducing, the spectrum touches the real axis. In other words, the real axis, which may be seen as the most natural candidate for a ``line gap'', does not separate the spectrum with respect to the two phases. While the more relevant criterion for a transition in our context, namely that the spectrum touches the real axis, is not considered in \cite{Kawabata2019symmetry}, it may be closer in spirit to the classification scheme discussed in \cite{Rudner2009topological}.

Our findings indicate that the focus on quantum trajectories, captured by the jumptime evolution, may be beneficial or possibly pivotal in characterizing the topology of the associated disspative quantum systems, beyond the considered frame, and irrespective of whether or not the trajectories are realized by monitoring and the quantum jumps detected. Similar to the Berry/Zak phase for closed quantum systems based on eigenstates, the jumptime phase based on jumptime propagators (with respect to invariant intracell states) can serve as a platform for a general topological classification beyond the considered cases, such as higher-dimensional intracell spaces and/or momentum-dependent jump operators.

Our results demonstrate that, in jump time, the bulk transport can serve as a reliable topology witness with potential for applications: If realized under monitoring, one may conceive a topologically controlled switching between different transport behaviors that takes effect from one quantum jump to the next. The questions of whether and how our topological classification is in addition reflected by a bulk-edge correspondence we leave for future research.

\paragraph*{Acknowledgments.}
We thank M. Cirio for helpful discussions.
F.N. is supported in part by:
Nippon Telegraph and Telephone Corporation (NTT) Research,
the Japan Science and Technology Agency (JST) [via
the Quantum Leap Flagship Program (Q-LEAP),
Moonshot R\&D Grant No. JPMJMS2061],
the Japan Society for the Promotion of Science (JSPS)
[via Grants-in-Aid for Scientific Research (KAKENHI) Grant No. JP20H00134],
the Army Research Office (ARO) (Grant No. W911NF-18-1-0358),
the Asian Office of Aerospace Research and Development (AOARD) (via Grant No. FA2386-20-1-4069), and
the Foundational Questions Institute Fund (FQXi) via Grant No. FQXi-IAF19-06.

\appendix

\section{}

Here, we describe the jumptime evolution for the Hamiltonian (\ref{Eq:Bloch_Hamiltonian}) in the most general case of $h_z(p) \neq 0$, and the collective collapse operator (\ref{Eq:collective_collapse_operator}).

The jumptime evolution, restricted to the $A$ sublattice, reads $\bra{p} \rho_{n+1}^{(A)} \ket{p'} = K_{\rm cc}(p,p') \bra{p} \rho_{n}^{(A)} \ket{p'}$, with the jumptime propagator
\begin{subequations} \label{Eq:general_collective_collapse_propagator}
\begin{align}
K_{\rm cc}(p,p') &= \int_{0}^{\infty} \!\!\! \gamma d \tau \bra{B} e^{-\frac{i}{\hbar} \hat{H}_{\rm eff}(p) \tau} \ket{A} \bra{A} e^{\frac{i}{\hbar} \hat{H}_{\rm eff}^{\dagger}(p') \tau} \ket{B} \nonumber \\
&= \frac{2 \hbar^2 \gamma^2 \big( h_x(p) + i h_y(p) \big) \big( h_x(p') - i h_y(p') \big)}{2 A^2(p,p') + \hbar^2 \gamma^2 B(p,p')} ,
\end{align}
where we abbreviated
\begin{align}
A(p,p') =& h_\perp^2(p) - h_\perp^2(p') + h_z^2(p) - h_z^2(p') \\
 &+ i \hbar \gamma \big[ h_z(p) + h_z(p') \big]/2 , \nonumber \\
B(p,p') =& h_\perp^2(p) + h_\perp^2(p') + h_z^2(p) + h_z^2(p') \\
 &+ i \hbar \gamma \big[ h_z(p) - h_z(p') \big]/2. \nonumber
\end{align}
\end{subequations}
As above, we require $h_\perp^2(p) = h_x^2(p) + h_y^2(p) \neq 0 \, \forall p$. Note that here again $K_{\rm cc}(p,p) = 1$, in agreement with normalization. When $h_z=0$, as expected, Eq.~(\ref{Eq:collective_collapse_propagator}) is recovered. To obtain (\ref{Eq:general_collective_collapse_propagator}), we write $\hat{H}_{\rm eff}(p) = -i \hbar \frac{\gamma}{4} \mathbb{1}_2 + \vec{h}_{\rm eff}(p) \cdot \vec{\sigma}$, with $\vec{h}_{\rm eff}(p) = \big( h_x(p), h_y(p), h_z(p) + i \hbar \frac{\gamma}{4} \big)^T$, and use
\begin{align}
\bra{B} & e^{-\frac{i}{\hbar} \vec{h}_{\rm eff}(p) \cdot \vec{\sigma} \tau} \ket{A} \nonumber \\
& = -i \frac{\tau}{\hbar} \sinc \big( \frac{\tau}{\hbar} h_{\rm eff}(p) \big) \big( h_x(p) + i h_y(p) \big) ,
\end{align}
where $h_{\rm eff}^2(p) = h_\perp^2(p) + h_z^2(p) - \frac{\hbar^2 \gamma^2}{16} + i \hbar \frac{\gamma}{2} h_z(p)$ and $\sinc(z) = \frac{\sin(z)}{z}$.

\section{}

For a generic Bloch Hamiltonian
$\hat H (p)$,
the jumptime phase
$T_{\rm cc}$
has nontopological contributions
$R_{1,2}$
defined by
Eqs.~(\ref{Eq:rest_terms}).
Here, we demonstrate that, if the Bloch Hamiltonian satisfies certain
symmetries, the non-topological terms vanish identically. 
	
We start with time-reversal symmetry
$\mathcal{T}$.
Under the action of $\mathcal{T}$, momentum changes sign,
$p \rightarrow -p$,
and the Hamiltonian is subject to complex conjugation. Consequently, the
Hamiltonian 
$\hat{H}$
is said to possess time-reversal symmetry when the following holds true:
\begin{eqnarray}
	\mathcal{T}: \quad \hat{H} (p) = \hat{H}^* (-p)
	\quad
	\forall p.
\end{eqnarray}
This implies that
\begin{subequations}
	\begin{align}
		h_x (p) &= h_x (-p),
		\\
		h_y (p) &= - h_y (-p),
		\\
		h_z (p) &= h_z (-p).
	\end{align}
\end{subequations}
In other words, 
$h_{x,z} (p)$
must be even functions of $p$, while
$h_{y} (p)$
is odd, which results in vanishing
$R_{1,2}$.
Indeed, it is easy to check that
$h_\perp^2$
is even, while
$\partial h_z/\partial p$
is odd. Thus, the integrands in
Eqs.~(\ref{Eq:rest_terms})
are odd, and the integrals are both zero.

Spatial inversion symmetry $\mathcal{P}$, on the other hand, does not
guarantee the nullification of the nontopological terms. Under the action
of $\mathcal{P}$, momentum changes sign,
$p \rightarrow -p$,
and the sublattices switch,
$A \leftrightarrow B$.
Therefore the system is invariant relative to the spatial inversion when
its Hamiltonian satisfies
\begin{eqnarray}
	\mathcal{P}: \quad \hat{H} (p) = \sigma_x \hat{H} (-p) \sigma_x \quad \forall p ,
\end{eqnarray}
which is equivalent to
\begin{subequations}
	\begin{align}
		h_x (p) &= h_x (-p),
		\\
		h_y (p) &= - h_y (-p),
		\\
		h_z (p) &= - h_z (-p).
	\end{align}
\end{subequations}
These relations are insufficient to argue that
$R_{1,2}$
vanish. Indeed, they imply that the integrands in
Eqs.~(\ref{Eq:rest_terms})
are even, and we cannot claim that
$R_{1,2} = 0$.
	
Next, we consider $\mathcal{PT}$ symmetry (invariance after simultaneous
inversion of both the spatial coordinate and the time direction). We demand
that
\begin{eqnarray}
	\mathcal{PT}: \quad \hat{H}(p) = \sigma_x \hat{H}^*(p) \sigma_x \quad \forall p .
\end{eqnarray}
A Hamiltonian satisfies this equality when
\begin{eqnarray}
	h_z (p) = - h_z (p) ,
\end{eqnarray} 
which further implies
\begin{eqnarray} 
	\label{eq::PT_symm}
	h_z(p) \equiv 0.
\end{eqnarray} 
Identical nullification of
$h_z$
guarantees that both non-topological terms vanish.
	
Finally, we discuss chiral symmetry. A Hamiltonian
$\hat{H}$
possesses chiral symmetry 
${\cal S}$
if 
\begin{eqnarray}
	{\cal S}: \quad \sigma_z\hat{H}(p)\sigma_z = -\hat{H}(p) \quad \forall p.
\end{eqnarray}
Speaking informally, a Hamiltonian satisfies this requirement when
intrasublattice terms are absent and only intersublattice hopping terms are
present. It is easy to check that 
Eq.~(\ref{eq::PT_symm})
is both necessary and sufficient for
$\hat{H}$
to exhibit chiral symmetry. This means that any chirally symmetric
Hamiltonian is also
${\cal PT}$ symmetric,
and the reverse is also true. Consequently, the nontopological terms are absent for a chiral Hamiltonian.

\section{}

We describe the jumptime evolution for the Hamiltonian (\ref{Eq:Bloch_Hamiltonian}) with $h_\perp^2(p) = h_x^2(p) + h_y^2(p) \neq 0 \, \forall p$, complemented by the sublattice projector $\hat{L}_{A} = \mathbb{1}_{\infty} \otimes \ket{A}\bra{A}$. For simplicity, we specify to $h_z(p) \equiv 0$.

The jumptime evolution, restricted to the $A$ sublattice, reads $\bra{p} \rho_{n+1}^{(A)} \ket{p'} = K_{A}(p,p') \bra{p} \rho_{n}^{(A)} \ket{p'}$, with the jumptime propagator
\begin{align} \label{Eq:sublattice_propagator}
K_{A}(p,p') &= \int_{0}^{\infty} \!\!\! \gamma d \tau \bra{A} e^{-\frac{i}{\hbar} \hat{H}_{\rm eff}(p) \tau} \ket{A} \bra{A} e^{\frac{i}{\hbar} \hat{H}_{\rm eff}^{\dagger}(p') \tau} \ket{A} \nonumber \\
&= \frac{\hbar^2 \gamma^2 \big( h_\perp^2(p) + h_\perp^2(p') \big)}{2 \big( h_\perp^2(p) - h_\perp^2(p') \big)^2 + \hbar^2 \gamma^2 \big( h_\perp^2(p) + h_\perp^2(p') \big)} .
\end{align}
Note that $K_A(p,p) = 1$, in agreement with normalization. The respective jumptime propagator for the sublattice projector $\hat{L}_{B} = \mathbb{1}_{\infty} \otimes \ket{B}\bra{B}$ reads identically, $\bra{p} \rho_{n+1}^{(B)} \ket{p'} = K_{B}(p,p') \bra{p} \rho_{n}^{(B)} \ket{p'}$ with $K_{B}(p,p')$ as in (\ref{Eq:sublattice_propagator}). To obtain (\ref{Eq:sublattice_propagator}), we write $\hat{H}_{\rm eff}(p) = -i \hbar \frac{\gamma}{4} \mathbb{1}_2 + h_x(p) \sigma_x + h_y(p) \sigma_y - i \hbar \frac{\gamma}{4} \sigma_z$ and use $\exp(-i \vec{a} \vec{\sigma}) = \cos(a) \mathbb{1}_2 - i \, \sinc(a) \, \vec{a} \cdot \vec{\sigma}$, where $\sinc(x) = \frac{\sin(x)}{x}$ and $a = \sqrt{a_x^2+a_y^2+a_z^2}$.

\section{}

We determine the steady state of the master equation (\ref{Eq:Lindblad_master_equation}) for the Hamiltonian (\ref{Eq:Bloch_Hamiltonian}) and the collective collapse operator (\ref{Eq:collective_collapse_operator}), and evaluate its transport behavior. For simplicity, we restrict ourselves to the case $h_z(p) \equiv 0$.

Since both $\hat{H}$ and $\hat{L}_{\rm cc}$ are diagonal in the momentum basis, the steady state $\rho_{\rm ss}$ of the master equation (\ref{Eq:Lindblad_master_equation}) is easily obtained in momentum representation, $\rho_{\rm ss} = \oint d p \, \bra{p} \rho_{{\rm ex}, 0} \ket{p} \ket{p}\bra{p} \otimes \rho_{\rm in}(p)$, where $\bra{p} \rho_{{\rm ex}, 0} \ket{p}$ denotes the momentum distribution of the initial state. In Bloch representation, the internal state component $\rho_{\rm in}(p) = \frac{1}{2} \left( \mathbb{1}_2 + \vec{r}_{\rm ss}(p) \cdot \vec{\sigma} \right)$ reads
\begin{align}
\vec{r}_{\rm ss}(p) = \frac{\hbar \gamma/2}{h_\perp^2(p) + \hbar^2 \gamma^2/8}
\left( \begin{array}{c}
h_y(p) \\ -h_x(p) \\ \hbar \gamma/4
\end{array} \right) .
\end{align}

To evaluate the transport behavior, we further specify to the SSH model and determine the expectation value of the single-particle cross-section current $\hat{J} = -i w (\ket{j+1, B}\bra{j, A} - \ket{j, A}\bra{j+1, B})$. Note that the steady state is translation invariant and therefore $\langle \hat{J} \rangle_{\rm ss}$ is independent of $j$. Assuming a localized initial state, $\rho_{{\rm ex}, 0} = \ket{j_0}\bra{j_0}$ or $\bra{p} \rho_{{\rm ex}, 0} \ket{p} = a/h$, we obtain
\begin{align}
\langle \hat{J} \rangle_{\rm ss} = \frac{\gamma \hbar}{4} \left( 1 + \frac{w^2-v^2 - \gamma^2 \hbar^2/8}{\big[ (w^2-v^2 - \gamma^2 \hbar^2/8)^2 + w^2 \gamma^2 \hbar^2/2 \big]^{1/2}} \right) .
\end{align}
The current is finite at any parameter values, and exhibits a smooth crossover from $\langle \hat{J} \rangle_{\rm ss} = \gamma \hbar/2$ at $w \gg v$ to $\langle \hat{J} \rangle_{\rm ss} = 0$ at $v \gg w$. The width of the crossover is controlled by $\gamma$. Only in the limit $\gamma \rightarrow 0$ is the topological transition at $v=w$ recovered.

\section{}

We here give the real-space Hamiltonian corresponding to the Bloch Hamiltonian (\ref{Eq:torus}). They are related through the basis transformation $\bra{j_k} p_k \rangle = \sqrt{a_k/h} \exp\left(i j_k a_k p_k/\hbar\right)$, $k=1,2$. The Hamiltonian expressed in terms of the position basis reads
\begin{align} \label{Eq:2D_real-space_Hamiltonian}
\hat{H}^{(2{\rm D})} = u \sum_{j_1, j_2 \in \mathbb{Z}} &\ket{j_1, j_2} \bra{j_1, j_2} \otimes \sigma_x \\
+ v \sum_{j_1, j_2 \in \mathbb{Z}} \Big[ &\ket{j_1, j_2} \bra{j_1+1, j_2} \otimes \frac{1}{2} (\sigma_x + i \sigma_y) \nonumber \\
+&\ket{j_1, j_2} \bra{j_1-1, j_2} \otimes \frac{1}{2} (\sigma_x - i \sigma_y) \Big] \nonumber \\
+ w \sum_{j_1, j_2 \in \mathbb{Z}} \Big[ &\ket{j_1, j_2} \bra{j_1, j_2+1} \otimes (\sigma_z + i \sigma_y) \nonumber \\
+&\ket{j_1, j_2} \bra{j_1, j_2-1} \otimes (\sigma_z - i \sigma_y) \Big] . \nonumber
\end{align}

The Hamiltonian (\ref{Eq:2D_real-space_Hamiltonian}) employs coordinates that are based on the primitive translation vectors $\{\vec{a}_1, \vec{a}_2\}$, cf.~Fig.~\ref{Fig:two_dimensions}. If we alternatively use the primitive translation vectors $\vec{a}'_1 = \vec{a}_1$ and $\vec{a}'_2 = \vec{a}_2 - \vec{a}_1$, we obtain the Hamiltonian
\begin{align}
\hat{H}'^{(2{\rm D})} = u \sum_{j_1, j_2 \in \mathbb{Z}} &\ket{j_1, j_2} \bra{j_1, j_2} \otimes \sigma_x \\
+ v \sum_{j_1, j_2 \in \mathbb{Z}} \Big[ &\ket{j_1, j_2} \bra{j_1+1, j_2} \otimes \frac{1}{2} (\sigma_x + i \sigma_y) \nonumber \\
+&\ket{j_1, j_2} \bra{j_1-1, j_2} \otimes \frac{1}{2} (\sigma_x - i \sigma_y) \Big] \nonumber \\
+ w \sum_{j_1, j_2 \in \mathbb{Z}} \Big[ &\ket{j_1, j_2} \bra{j_1+1, j_2+1} \otimes (\sigma_z + i \sigma_y) \nonumber \\
+&\ket{j_1, j_2} \bra{j_1-1, j_2-1} \otimes (\sigma_z - i \sigma_y) \Big] , \nonumber
\end{align}
with the corresponding Bloch representation
\begin{subequations} \label{Eq:transformed_torus}
\begin{align}
h'_x(\vec{p}) &= u + v \cos \left( \frac{p_1 a_1}{\hbar} \right) , \\
h'_y(\vec{p}) &= v \sin \left( \frac{p_1 a_1}{\hbar} \right) + 2 w \sin \left( \frac{p_1 a_1 + p_2 a'_2}{\hbar} \right) , \\
h'_z(\vec{p}) &= 2 w \cos \left( \frac{p_1 a_1 + p_2 a'_2}{\hbar} \right) .
\end{align}
\end{subequations}
Comparing (\ref{Eq:torus}) and (\ref{Eq:transformed_torus}), we find the transformation $\vec{h}'(p_1, p_2) = \vec{h}(p_1, \frac{a'_2}{a_2} p_2 + \frac{a_1}{a_2} p_1)$.

\bibliography{references}

\end{document}